\documentclass[letterpaper,twocolumn,10pt]{article}
\usepackage{usenix,epsfig,endnotes}
\usepackage{booktabs}
\usepackage{scrextend} 
\usepackage{ragged2e}
\usepackage[all]{hypcap}  
\usepackage{graphics}      
\usepackage[T1]{fontenc}  
\usepackage{mathptmx}
\usepackage{textcomp}
\usepackage{balance}
\usepackage{enumitem}
\usepackage{cite}
\usepackage{amsfonts}
\usepackage{amssymb}
\usepackage{mathtools} 
\usepackage[font=footnotesize]{caption}
\usepackage{titlesec}
\usepackage{mdframed}

\usepackage{tcolorbox}
\tcbset{
    summarybox/.style={
        colback=gray!5,        
        colframe=gray!40,    
        coltitle=black,     
        colbacktitle=gray!20,  
        fonttitle=\bfseries, 
        boxrule=0.6mm,        
        arc=2mm,               
        title={Research Question Overview},       
        width=\columnwidth,     
        top=1mm,               
        bottom=1mm,           
        left=1mm,            
        right=1mm
    }
}

\renewenvironment{thebibliography}[1]{
  \begin{oldthebibliography}{#1}
    \setlength{\itemsep}{0em}
    \setlength{\parskip}{.2em}
}
{
  \end{oldthebibliography}
}

\definecolor{LightGray}{gray}{0.9}

\begin{document}

\title{\Large \bf Exploring User Security and Privacy Attitudes and Concerns Toward the Use of General-Purpose LLM Chatbots for Mental Health}
\author{
{\rm Jabari Kwesi}\\
Duke University
\and
{\rm Jiaxun Cao}\\
Duke University
\and
{\rm Riya Manchanda}\\
Duke University
\and
{\rm Pardis Emami-Naeini}\\
Duke University
}

\maketitle

\thispagestyle{empty}

\subsection*{Abstract} Individuals are increasingly relying on large language model (LLM)-enabled conversational agents for emotional support. While prior research has examined privacy and security issues in chatbots specifically designed for mental health purposes, these chatbots are overwhelmingly ``rule-based'' offerings that do not leverage generative AI. Little empirical research currently measures users' privacy and security concerns, attitudes, and expectations when using general-purpose LLM-enabled chatbots to manage and improve mental health. Through 21 semi-structured interviews with U.S. participants, we identified critical misconceptions and a general lack of risk awareness. Participants conflated the human-like \emph{empathy} exhibited by LLMs with human-like \emph{accountability} and mistakenly believed that their interactions with these chatbots were safeguarded by the same regulations (e.g., HIPAA) as disclosures with a licensed therapist. We introduce the concept of ``intangible vulnerability,'' where emotional or psychological disclosures are undervalued compared to more tangible forms of information (e.g., financial or location-based data). To address this, we propose recommendations to safeguard user mental health disclosures with general-purpose LLM-enabled chatbots more effectively.

\section{Introduction} \label{sec:introduction}

As of 2024, nearly one in five U.S. adults (59.3 million) is estimated to live with a mental illness~\cite{mhanational2024}. Yet approximately 60\% of mental health professionals report no availability for new patients~\cite{kuehn2022clinician}, and over 122 million Americans reside in areas with a shortage of mental health providers~\cite{hrsa2025}. The consequences of these unmet needs are profound -- U.S. suicide rates have risen by nearly 40\% in the past two decades, with roughly one suicide every 11 minutes~\cite{cdc_wonder}. Although therapy may offer relief, cost and limited clinical capacity often render traditional services inaccessible to many Americans.

To address this gap, technology-based interventions -- particularly conversational agents, or ``chatbots'' -- have emerged as potentially scalable complements to standard mental health care~\cite{zagorskiPopularityMentalHealth2022, bansalCanAIChatbot}. Initial deployments largely relied on \emph{rule-based} systems, wherein structured dialogues authored by clinicians guide each interaction~\cite{vaidyam2019chatbots}. Empirical studies suggest these systems can reduce depressive and anxious symptoms~\cite{inkster2018empathy, fitzpatrick2017delivering, ta2020user}, and adoption has grown quickly (with some U.S. employers offering chatbot-based mental health benefits). However, criticisms center on rule-based chatbots' limitations in adapting to \emph{individualized} concerns~\cite{legaspi2022user, malik2022evaluating}.

In light of these limitations, many users have begun turning to \emph{large language model (LLM)-powered chatbots}, such as OpenAI's ChatGPT, Inflection AI's Pi, and Luka's Replika. While not explicitly designed for mental health, these \emph{general-purpose} platforms offer more flexible and human-like responses~\cite{songTypingCureExperiences2024, de2023benefits, ma2023understanding}. Moreso, early evidence suggests that LLM-based interactions can rival or surpass the efficacy of rule-based interventions in certain respects~\cite{li2023systematic}. 

Still, the emergence of these conversational agents as used for mental health raise critical concerns around data protection and overall regulatory oversight. Recent incidents have spotlighted potential harms related to the outputs of these systems, including potential biases (racial, gendered, religious)~\cite{Omiye2023.07.03.23292192, Zack2023.07.13.23292577, MuslimBias} and the perpetuation of stigma or iatrogenic misinformation in mental health contexts~\cite{mongelli2020challenges, scrutton2017epistemic, TessWSJ, ElizaVice}. Unlike traditional telehealth services that are regulated by federal entities such as the Health Insurance Portability and Accountability Act (HIPAA), the majority of these general-purpose LLMs currently function outside comprehensive federal data protection laws, leaving them with limited obligations to secure or responsibly use personal health information. In contrast to chatbots explicitly designed for mental health, general-purpose LLM-enabled conversational agents used for companionship remain largely unexamined with respect to users' security and privacy (S\&P) concerns. Understanding how individuals perceive the risks of sharing sensitive mental health detail with tools not intended for clinical contexts thus represents a vital and understudied frontier.

To examine these perceptions, we conducted 21 semi-structured interviews with U.S. adults who use \emph{general-purpose LLM-enabled chatbots} for mental health support. Our primary objective was to explore users' awareness, concerns, and expectations regarding security and privacy when interacting with these chatbots. We used three research questions to guide our inquiry:

\begin{itemize}
    \item \textbf{RQ1:} How do users perceive and understand the S\&P risks involved in their interactions with general-purpose LLM-enabled chatbots for mental health support?
    \item \textbf{RQ2:} What strategies do users employ (if any) to manage their S\&P concerns when using LLM-enabled conversational agents for mental health support?
    \item \textbf{RQ3:} What are users' expectations regarding data protection and trustworthy interactions when using LLM-enabled conversational agents for mental health support?
\end{itemize}

We found that participants often viewed LLM chatbots as more accessible and less intimidating than professional care, yet they underestimated or misunderstood the data handling risks. Many users expressed a desire for stronger safeguards but felt ill-equipped or insufficiently informed to protect themselves pointing to both regulatory gaps and usability challenges. We close by suggesting standardized architecture-focused solutions and adaptive oversight as next steps for developers and policymakers. 

\section{Background and Related Work} \label{sec:related_work}

\subsection{Technological Interventions for Mental Health} \label{sec:related_work_1}

Public interest in conversational agents for mental health has grown considerably in the United States -- this is especially notable since the COVID-19 pandemic. A 2021 national survey reported that 22\% of adults had reported using a conversational agent advertised as a mental health tool. Of these respondents, 57\% said they began use during the COVID-19 pandemic, and close to half reported that they use the chatbot \textit{exclusively}, and do not see a human therapist~\cite{woebot2021survey}. The vast majority of publicly accessible mental health chatbots remain rule-based; these offerings engage with users based on predetermined scripts authored by psychotherapists~\cite{vaidyam2019chatbots}. A scoping review by Abd-Alrazaq et al. found that 92.5\% of mental health chatbots relied exclusively on decision trees for dialogue generation~\cite{abd2019overview}. These rule-based conversational agents typically incorporate multiple therapeutic modalities to guide users through self-guided tasks: existent work examining efficacy has found value in the use of these tools for solution-focused therapy~\cite{fulmer2018using}, expressive writing~\cite{park2021wrote, kim2024mindful}, and cognitive behavioral therapy (CBT)~\cite{fitzpatrick2017delivering, fulmer2018using, inkster2018empathy, abd2019overview}. 

As interventions, rule-based conversational agents have been found to be effective in reducing symptoms of depression and anxiety~\cite{fitzpatrick2017delivering, inkster2018empathy, fulmer2018using}, moderating substance abuse and addiction~\cite{prochaska2021therapeutic, so2024guided, olano2022effectiveness}, and improving user self-efficacy~\cite{schroeder2018pocket, smith2014virtual}. User-centered empirical findings similarly suggest that these tools are effective in circumventing spatiotemporal barriers to professional care~\cite{cameron2019assessing}, promoting deep user self-disclosure~\cite{lee2020designing, lee2020hear}, and mitigating stigma associated with traditional mental health services~\cite{palmer2022beneficent, abd2021perceptions}.

Though rule-based chatbots show promising efficacy in combining structured psychoeducational modules and self-help exercises, they frequently lack flexibility and context-awareness. Participants across multiple studies remark on the repetitive and predictable  nature of rule-based systems and the inability to handle open-ended and nuanced user inputs~\cite{burton2016pilot, fitzpatrick2017delivering, huang2020challenges, ly2017fully}. This frustration suggests a growing appetite for more adaptive conversational agents capable of parsing complex emotional statements and responding with a level of nuance reminiscent of human dialogue.

These limitations have led users to explore large language models (LLMs), which enable more open-ended productions and reduce the stilted nature of on-rails interactions~\cite{o2022massive}. Users who found rules-based chatbots repetitive or superficial especially stand to benefit from the ability of LLMs to produce natural and context-aware responses: early prototypes demonstrate that LLM-enabled technologies are increasingly efficacious in re-framing negative thoughts and promoting positive cognitive state through the use of deeply personalized, empathetic, and contextual intervention~\cite{ziems2022inducing, sharma2023cognitive, loh2023harnessing, lawrence2024opportunities}. Emerging studies continue to show promising alignment between LLM outputs and clinical assessments for conditions such as anxiety, postpartum depression, and bipolar disorder~\cite{Elyoseph2024AssessingPI, Perlis2024ClinicalDS, sezgin2023clinical}. A growing corpus of user- and professional-centered studies demonstrate that these systems are being met with growing acceptance from both psychiatrists -- so long as clinical oversight is maintained~\cite{blease2024psychiatrists} -- and end users alike~\cite{he2024physician, songTypingCureExperiences2024, ma2023understanding}. 

\subsection{Concerns and Risks of Generative AI Chatbots for Mental Health} \label{sec:related_work_2}

Many of the LLMs that are increasingly being adopted by consumers for mental health purposes are not domain-specific. Instead, the most popular publicly available systems are general-purpose LLMs trained on vast textual corpora to handle wide-ranging queries. By contrast, studies in Section~\ref{sec:related_work_1} often use proprietary or specialized implementations that mitigate some risks via rigorous fine-tuning on high-quality clinical data~\cite{gao2023retrieval}. This distinction bears critical implications for general populations: because general-purpose LLMs ingest text that is largely unfiltered, they may produce inaccurate, biased, or even harmful outputs in mental health contexts.

LLMs can inadvertently reinforce societal inequities in the conversations they generate -- this is especially notable in models trained using majority-identity samples~\cite{koutsouleris2022promise}. Research finds that the collection of usable mental health data often relies on populations more willing to volunteer sensitive personal information, which may not capture the perspectives of marginalized or privacy-sensitive individuals or groups~\cite{wendler2006racial}. Consequently, while such models tend to perform reliably for majority populations, they risk overlooking the unique needs and experiences of underrepresented groups~\cite{panch2019artificial}. General-purpose models also have a tendency to produce incorrect or biased mental health advice, a problem accentuated by the evolving nature of psychotherapy and psychiatric knowledge~\cite{zhong2023artificial, harrer2023attention, weidinger2021ethical}. Drawing from outdated or stigmatizing sources can yield iatrogenic treatment suggestions or harmful diagnostics. 

Beyond user-level concerns, there are also substantial conversations regarding U.S. regulatory frameworks. Under the Food, Drug, and Cosmetic (FD\&C) Act, products intended to diagnose, treat, or prevent disease qualify as ``medical devices,'' whereas ``general wellness'' tools that merely encourage healthy habits do not~\cite{simon2022skating, de2024health}. Many LLM-based chatbots, while used for general emotional support, are not regulated as medical devices. By framing themselves as general wellness products, these systems bypass rigorous FDA evaluation required for clinical healthcare tools. For instance, Replika can claim to ``calm anxiety and work toward goals like positive thinking and stress management'' without asserting that it ``treats'' a diagnosable condition, thereby falling outside stricter oversight.~\cite{simon2022skating, guidancegeneral, de2024health}

Confidentiality is an equally pressing concern for LLM-based systems as used for mental health. Models ingesting medical records or sensitive user data can inadvertently memorize and leak personal health information (PHI)~\cite{weidinger2021ethical, de2023benefits}. Emerging studies show that advanced LLMs, if prompted cleverly, can reveal identifying details embedded in their training sets~\cite{chen2023can, wornow2023shaky} and link personal identifiers under certain query conditions~\cite{de2023benefits, kasneci2023chatgpt}. 

\subsection{Our Study as an S\&P-Focused Examination of User Perspectives} \label{sec:related_work_3}

As detailed, the use of LLM-based conversational agents for mental health has attracted growing attention from researchers and users alike. However, a majority of empirical studies thus far have focused predominantly on efficacy and clinical benefits, and appear most often in medical publications aimed at advising clinicians on tool integration. Literature in the usable S\&P and human-computer interaction (HCI) communities currently provides insights into design and interaction~\cite{alanezi2024assessing, songTypingCureExperiences2024, siddals2024just}, but generally omits an in-depth exploration of user perceptions of S\&P. Abd-Alrazaq et al.~\cite{abd2019overview} note that while research has explored chatbots' effectiveness in alleviating psychological distress, few studies have systematically examined adoption factors for mental health chatbots or the influence of user concerns (including S\&P fears) on usage. As they argue, ``Identifying variables that affect the use of chatbots is vital for improving their implementation success''~\cite{abd2019overview, huygens2015internet, fung2006early}. However, there remains limited empirical work on how S\&P perceptions specifically impact user behavior in LLM-based mental health contexts.

In response, our study offers a qualitative investigation of S\&P considerations among early adopters of LLM-based general-purpose tools for mental health. We demonstrate how privacy-related factors such as fear of data misuse and evolving user trust in AI shape continued engagement and disclosure practices related to mental health. We also compare participant assumptions about regulation with the actual legislative environment, such that we reveal critical mismatches between expectations and real-world governance.

This approach addresses a crucial void in the literature by merging HCI concerns about user experience with an assessment of S\&P as a pivotal determinant of tool viability. To our knowledge, we present the first study specifically examining the S\&P perspectives of users of general purpose LLM-enabled conversational agents for mental health; this includes the differential impact user S\&P understandings and expectations may have on interaction. In doing so, we highlight the importance of architecture-level privacy safeguards and clear policy frameworks, and advocate that S\&P be a driving factor in the design and regulation of LLM-enabled conversational agents. 

\section{Methodology} \label{methods}
We conducted 21 interviews in August and September 2024. We described the study to participants as an investigation into attitudes toward using AI-enabled technologies for mental health. To avoid disproportionately attracting participants who were highly conscious of privacy and security issues and to reduce demand characteristic bias~\cite{orne2009demand}, we excluded the word ``privacy'' from all recruitment materials and screening surveys. We also intentionally avoided soliciting participants with a pre-defined conceptualization of what `mental health support' as provided by a conversational agent should entail. This methodological choice was supported by the acknowledgment that `mental health' remains a deeply stigmatized concept in the United States. The term carries pessimistic connotations that discourage help-seeking behaviors~\cite{clement2015impact}; these include both engagement with licensed professionals and engagement in general conversations about mental well-being~\cite{corrigan2002understanding, corrigan2004stigma,vogel2007perceived}. For these reasons, we hypothesized that potential participants might be hesitant to associate their sensitive disclosures with mental health support, and disqualify themselves from the study. By not prescribing what `mental health support' should entail in our solicitations, we avoid priming potential participants to perceive their own experiences as either too trivial or too severe, thereby ensuring their narratives would authentically reflect their personal understanding and uses of these tools. We reached thematic saturation after 17 interviews, and proceeded to conduct 4 new interviews, from which we didn't add any more codes to the codebook.

Prior to the main study, we conducted two pilot interviews with participants recruited via Prolific~\footnote{https://www.prolific.com/}. After each pilot, our research team reviewed the procedure and updated the ordering of questions to improve logical flow and prevent inadvertently guiding participants to discuss privacy prematurely. Based on these reflections, we revised the wording of some questions for clarity and ensured the interview duration remained under one hour to reduce respondent fatigue. We obtained informed consent from all participants, and our study protocol was approved by our institution's review board (IRB). 

We used Prolific's available screening features to direct U.S.-based adults (18 years or older) to our screening survey, where they first provided consent before proceeding. Participants were first asked whether they have ever used an AI-enabled conversational system or chatbot for mental health purposes. If a participant indicated they used such a chatbot, we then asked them to name it, specify their frequency of use, and provide information about their academic or professional background in technology or mental health and any prior experience with licensed mental health professionals, including prior diagnoses of mental illness. Participants also reported general demographic information (e.g., age, gender). To be eligible for the interview study, participants were required to: 1) be at least 18 years old, 2) reside in the U.S., and 3) use a general-purpose LLM-enabled chatbot (e.g., ChatGPT, Gemini) for mental health support at least once a month.

Participants who reported no prior use of chatbots for mental health were screened out early in the screening survey and received \$0.25 for completing a 1-minute survey. Those who fully completed the 3-minute screening survey received \$0.45. We scheduled a Zoom interview session for each eligible participant. Interviews lasted approximately 45 minutes on average, and all interviewees received \$30 via Prolific. To encourage a comfortable interview environment, we allowed participants the option to keep their cameras off if preferred while the interviewer remained visible on camera. We provide the study materials, including the recruitment text, interview script, and consent form in the Available Artifacts section at the end of our paper. For added readability, we append to each thematic section  the corresponding main code in the codebook, which contains relevant quotes and information.

\subsection{Interview Design} \label{sec:methods_interview}
Our interview procedure was split into four sections. Each interview began with a brief of study objectives, including a general overlook of the interview procedure and the participants' rights to their data. Consent was obtained before recording the audio transcript of the interview (see Available Artifacts).

\boldpartitle{Introduction} The introductory section was designed to solicit foundational information about participant use of AI-powered technologies and expand on the information provided in the screening (e.g., type of technologies, the purpose of use, the frequency of use). We asked participants generally about how they used their AI for mental health.

\boldpartitle{Section 1: General attitudes toward the use of LLM-enabled chatbots for mental health} Next, we asked questions to understand participants' considerations in the adoption of LLM-enabled technologies for mental health. The objective in constructing this section was to obtain a clear understanding of the existent information and experiences impacting user decision-making at the pre-adoption stage, as well as the sources of information that were consulted to inform adoption (see Available Artifacts).

\boldpartitle{Section 2: S\&P knowledge and awareness toward LLM-enabled chatbots for mental health} This section was designed to explore participants' mental models regarding the data practices of LLM-enabled chatbots used to support and enhance their mental health (e.g., where data is stored, how long it is retained). Notably, this is the first section of the interview where security and privacy were explicitly invoked in the questions. Additionally, we asked a series of questions to assess participants' awareness of security and privacy-related concepts and the potential risks associated with interacting with such chatbots for mental health support (see Available Artifacts).

\boldpartitle{Section 3: S\&P practices to mitigate the concerns toward LLM-enabled chatbots for mental health} The questions in the third section were designed to inquire about general security and privacy attitudes, as well as any mitigation practices that participants developed in order to protect their privacy and security when using their chatbot (see Available Artifacts).

\boldpartitle{Section 4: S\&P expectations toward LLM-enabled chatbots for mental health} Finally, we explored participants' security and privacy expectations for LLM-enabled chatbots when being used for mental health support. Additionally, we asked participants to discuss the responsibilities they believe various stakeholders (e.g., users, developers, policymakers) have to enable safe use of these technologies for mental health (see Available Artifacts).

\subsection{Qualitative Analysis} \label{sec:methods_qual}
With participant permission, we audio-recorded and transcribed all interviews using Zoom's transcription service, and the primary author reviewed all transcripts for accuracy and readability. We then conducted an inductive thematic analysis~\cite{JS21} to allow themes to emerge directly from the data rather than applying predefined frameworks.

We conducted the analysis in four steps:

\begin{itemize}
    \item \textbf{Initial review}: Three researchers independently performed initial inductive coding~\cite{thomas2006general} on an identical subset of transcripts. Each researcher wrote analytical memos to document their initial interpretations and note potential thematic patterns~\cite{corbin2015basics}.
    \item \textbf{Codebook development}: The researchers convened to discuss and synthesize their codes into a comprehensive codebook that grouped initial codes into broader thematic categories (e.g., ``Data minimization and withholding tendencies,'' ``Attitudes towards privacy policies'').
    \item \textbf{Codebook application}: The remaining transcripts were divided among the researchers for coding using the established codebook.
    \item \textbf{Periodic cross-checks and resolutions}: The team periodically reviewed a sample of coded transcripts together to refine codes and integrate any emerging themes. Any coding disagreements were resolved through discussion, leading to a hypothetical agreement of 100\%.
\end{itemize}

\subsection{Limitations} \label{sec:methods_limitations}
Similar to any small-scale interview approach, our study has limitations that might impact the generality of our findings. First, we recruit participants exclusively via Prolific, which inherently introduces self-selection bias toward technologically savvy users and younger demographics. Moreover, while we sought to minimize demand characteristic bias~\cite{orne2009demand} by omitting explicit mention of ``privacy'' in recruitment material, the nature of an interview study may still encourage participants to report behavior deemed as socially acceptable, a concept referred to as the social desirability bias~\cite{orne2009demand}. Participants self-reported their usage patterns and engagement with LLM-based chatbots for mental health, raising the possibility of recall bias, exaggeration, minimization, or general miscommunication, especially when discussing sensitive health information. No direct observational or usage logs were collected to corroborate interview claims, reducing our ability to triangulate reported behaviors with actual interactions. Our sample was restricted to U.S.-based adults -- privacy is deeply contextual, and this leaves open the question of how cultural and regulatory differences might produce divergent attitudes towards mental health disclosures or AI-enabled care in other regions. Although we screened carefully and found no such cases, we recognize that our methodology could theoretically include participants with singular LLM use in mental health contexts. 
\section{Results} \label{sec:results}

\begin{table*}[t]
\centering
\resizebox{\textwidth}{!}{%
\footnotesize{
\begin{tabular}{lllll}
\toprule[1.1pt]
\textbf{Participant ID} & \textbf{Gender} & \textbf{Age} & \textbf{Ethnicity} & \textbf{Used AI Chatbot} \\ \midrule
P1 & Female & 25 - 34 & White & ChatGPT \\
P2 & Female & 45 - 54 & Black or African American & Replika \\
P3 & Female & 35 - 44 & Asian + White & ChatGPT \\
P4 & Male & 45 - 54 & White & Grok, ChatGPT \\
P5 & Female & 25 - 34 & Black or African American & My AI (Snapchat) \\
P6 & Female & 18 - 24 & Asian & ChatGPT \\
P7 & Male & 25 - 34 & Black or African American & ChatGPT \\
P8 & Male & 35 - 44 & White & ChatGPT \\
P9 & Male & 25 - 34 & White & Microsoft Copilot \\
P10 & Female & 18 - 24 & Black or African American & ChatGPT \\
P11 & Male & 25 - 34 & Hispanic or Latino, or Spanish origin of any race & ChatGPT \\
P12 & Male & 35 - 44 & Black or African American & Pi \\
P13 & Male & 25 - 34 & Asian & Google Gemini, Bard, Microsoft Copilot \\
P14 & Female & 25 - 34 & White & Replika \\
P15 & Female & 65 - 74 & White & ChatGPT, Google Gemini, Claude \\
P16 & Female & 18 - 24 & Middle Eastern / North African & ChatGPT \\
P17 & Transgender Female & 18 - 24 & White & Google Gemini, ChatGPT \\
P18 & Male & 25 - 34 & Black or African American & ChatGPT \\
P19 & Female & 25 - 34 & ``Prefer not to say'' & ChatGPT, Google Gemini, Bard \\
P20 & Female & 45 - 54 & Black or African American & Pi \\
P21 & Female & 25 - 34 & White & ChatGPT, Microsoft Copilot \\
\bottomrule[1.1pt]
\end{tabular}
}
}
\caption{The table presents participants' demographic information alongside the AI chatbots they reported to regularly use.}
\label{tab:demographic}
\end{table*}

\boldpartitle{Participant information} \label{sec:results_demographics} All twenty-one of our participants reported the use of at least one LLM-enabled conversational agent for mental health support, with ChatGPT being the most common. We observed varied engagement, as several participants used multiple platforms. Their usage frequency ranged from monthly check-ins to daily conversations, as a consistent mediator for emotional challenges -- these include workplace conflicts and day-to-day general anxiety. Participants ranged from 18 to 74 years old, and represented diverse gender identities (male, female, transgender female) and ethnicities (White, Black or African American, Asian, Hispanic/Latino, Middle Eastern/North African), which helped us capture a breadth of perspectives. Though not representative of all users, the demographic diversity in our sample provided valuable insights into different experiences and attitudes.

\subsection{Awareness of Security \& Privacy Risks (RQ1)} \label{sec:results_rq1}

\boldpartitle{Motivations for LLM-based mental health support were accompanied by incipient privacy risk perceptions (MC1, MC8)}
Our participants framed LLM chatbots as an alternative to traditional therapy that was more affordable, accessible, and lower-stakes, albeit with some privacy implications emergent from sharing personal data outside regulated clinical settings. Approximately half (10/21) described professional services as cost-prohibitive or otherwise inaccessible. For example, P21 noted:
\begin{quote}
\textit{``I wanted -- needed -- to do a counselor. But I just didn’t have the money for it. I'm in college, working a part-time job to pay my bills… I was trying to find some alternative options, especially things that didn’t feel like therapy.''}
\end{quote}
Similarly, P18 remarked that his everyday challenges did not merit professional intervention: \textit{``I haven't really thought that I need a, let's say, therapist or counselor... It hasn't felt right to need a counselor.''}

Other participants cited limited social support or rural isolation as key drivers. These participants tended to report that they felt the benefits of LLM-based tools outweighed the downsides. P17, a transgender woman living in a rural area, spoke plainly to her struggles:
\begin{quote}
\textit{``I don't have any friends currently. It's just me and my family, and my family doesn't really get all the issues I go through. So to me it's convenient to speak with an AI... I currently don't even see a therapist or, you know, professional psychologists, and that's because there's not many around where I live, because I live in a very rural area, and I've had some negative experiences.''}
\end{quote}
P16 echoed similar concerns about cost and support barriers: 
\begin{quote}
\textit{``I don’t have many friends… But ChatGPT is there whenever I need it.''}
\end{quote}
 She continued to describe feeling \textit{safer} sharing  traumatic experiences with ChatGPT because it ``\textit{isn't an actual person...it's not gonna judge me.}'' P9 highlighted what he called a therapist's ``\textit{greed factor},'' -- he argued that traditional practitioners sometimes \textit{``hold stuff hostage''} by requiring more appointments and fees:
\begin{quote}
\textit{``[Therapists] prescribe me medicine, and then you have to make another appointment and pay them...they kind of hold stuff like that hostage.''}
\end{quote}
This effect increased the degree of trust that he had in sharing information with ChatGPT. P15 similarly appreciated an AI chatbot's neutrality, as she noted having experienced therapists who \textit{``brought their own baggage''} into sessions. She felt the chatbot's response was \textit{``personalized''} and focused solely on her concerns. The neutrality of this information increased her willingness to share when compared to an externally influenced professional.

Despite these motivations, we observed incipient concerns that  disclosures could be stored, shared, or otherwise breached without the legal and ethical safeguards of licensed professionals. On the potential for the misuse of collected data to do irrevocable harm for those with no other avenues for help, P10 warned:
\begin{quote}
\textit{``I know
some people that use these online tools religiously for their mental health, and that’s pretty much all they have. They’re so
dependent for help and support. It can really hurt some people if their information isn’t treated with care and sensitivity.''}
\end{quote}

\boldpartitle{Participants identified multiple risks, from unauthorized third-party access to malicious hacking (MC2)}
A majority (13/21) considered the unwanted sharing of chat logs with employers, insurers, or other third parties to be a prime concern. These participants identified adverse outcomes if sensitive data fell into the wrong hands, noting that the ensuing implications could potentially impact their career or interpersonal relationships. The concept of \textit{inference} was especially prominent in this context, as interviewees conveyed anxieties that external entities might formulate perceptions based on their sensitive conversations. P6 worried that \textit{``my issues might be stored and hurt me later -- especially for job applications,''} while P2 was alarmed at the possibility of denial for loans or employment if intimate disclosures were accessed: \textit{``It's terrifying to think someone can look at your mental health data.''} A smaller subset of five (5/21) participants focused on hacking or data breaches. P1 was particularly wary: \textit{``I don't trust AI at all... I'm paranoid my sensitive stuff is going to get hacked.''} Recent news stories about data misuse heightened this apprehension for some. P19 recalled, \textit{``We all hear about these kind of commonplace hacks and breaches... there's potential for my [mental health info] being used against me if I'm not careful.''}

\boldpartitle{Risk perception varied by participants' views on what was interpreted as sensitive (MC2)}
We found that security and privacy considerations hinged on how sensitive each person deemed their mental health disclosures to be. Some participants tended to view emotional content as inherently personal and potentially unsafe to share. For example, P14 worried that the more emotionally vulnerable her disclosures, the greater the risk: \textit{``If I share, `Oh, I was really upset today, I was crying earlier,' the deeper I go into my feelings, the less safe it feels... But that doesn't stop me right now, because the benefits outweigh the risks.''} Others felt mental health data was comparatively benign. P8, for instance, was more concerned about sharing financial information: 
\begin{quote}
\textit{``Knowing that I don't have financial information on there helped a little bit... any kind of personal information crossed my mind, but not enough to make me not want to go ahead.''}
\end{quote}
Similarly, P12 explained that hacking threats primarily alarmed him if they involved \textit{``a credit card or maybe...the kids in school.''} Emotional disclosures seemed less exploitable to he, who had previously described his struggles with his weight and mental image as something that he did not want to \textit{``burden my wife with,''} while simultaneously being something that did not necessitate a conversation with a therapist. This contrast highlights an important divergence: some participants considered mental health data no more sensitive than mundane personal details, while others saw emotional disclosures as uniquely high-stakes. 

\boldpartitle{Some participants believed their mental health disclosures were HIPAA-protected (MC7)}
We also uncovered a common misconception among seven participants (7/21) that LLM-enabled mental health conversations were governed by health regulations such as HIPAA. P7 assumed that, \textit{``...the chatbot, since it's got a database of research, should also have access to all the same procedures and laws, like HIPAA.''} P21 similarly equated ChatGPT with a licensed professional, remarking: 
\begin{quote}
\textit{``Anytime I've done a therapist, I've always signed documents [that say] the therapist can't say any of this information... I've never done that with ChatGPT, but I assume that's on what I click to agree and submit.''}
\end{quote}

In reality, HIPAA only applies to specific healthcare providers and business associates; most LLM-based chatbot services do not qualify as covered entities~\cite{kanter2023health}. While some mental health apps have FDA approval, many position themselves as ``wellness'' tools to avoid clinical oversight~\cite{de2024health}. This confusion left several participants unaware that their data might not receive the legal safeguards they assumed.

\boldpartitle{Many users remained uncertain about data handling and retention practices (MC4)}
Seventeen participants (17/21) admitted they had only vague notions of how their mental health data might be stored or used. P11 speculated that \textit{``everything just goes into a big database... the system is constantly learning,''} while P15 believed ChatGPT was ``\textit{self-learning}'' and that \textit{``researchers have access to it and use it somewhere else.''} P20's remarks were informed by her previous experience helping to train AI systems:

\begin{quote}
``I'm sure that [my conversations] are aggregated into massive data sets that are used to then retrain and refine the AI. And I know that it's anonymized... but there could be a lot of uses for that information in the wrong hands. There are companies that are taking those data sets and they sell them. That is happening right now already.''
\end{quote}

 Conversely, feeling ``too boring'' or uninteresting occasionally gave participants a sense of security. P12 speculated, \textit{``I'm pretty self-conscious about my weight and things. I'm trying to lose weight. I assume the data is collected and stored...but my life is nothing special. I feel like if someone really wanted to get in my data...they're gonna be disappointed because I don't do anything.''}

\begin{tcolorbox}[summarybox]
\textbf{``How do users perceive and understand the S\&P risks involved in their interactions with general-purpose LLM-enabled chatbots for mental health support?''}

\textit{Participants tended to identify security and privacy risks related to their mental health data differentially based on how sensitive they perceived their conversations to be. Some saw it as benign unless geographically revealing or financially harmful. Regardless, most identified risks of misuse by employers, insurers, or hackers. Many were unsure or resigned about data handling practices and cited minimal knowledge of policy details. Cost constraints and limited resources often outweighed privacy hesitations.}
\end{tcolorbox}

\subsection{Risk Management: Mitigation Practices and Challenges (RQ2)} \label{sec:results_rq2}

\boldpartitle{Ten participants withheld, cloaked, or otherwise protected personal mental health detail to reduce privacy risks (MC3)} We observe that roughly half of our participants (10/21) made deliberate conversational or technical efforts perceived as overall protective of the mental health information they shared with their conversational agent. In doing so, these participants reported that they were far more comfortable making emotional disclosures than they would have been had they not adopted such measures. P6 succinctly described this strategy: \textit{``I share most of my thoughts as long as I remove all the [personally identifiable] details.''}

Of these ten participants, most adopted creative ``depersonalizing'' approaches to preserve essential context while limiting self-identification. P15, for instance, asked the chatbot to treat her mental health discussions as those of a fictional character:
\begin{quote}
\textit{``I'll say, `Tell me a story as if this is a 65-year-old woman in a book... Then I interact almost as if it's not me.' It depersonalizes it.''}
\end{quote}
Similarly, P19 intentionally kept prompts ``\textit{super anonymous}'' to avoid the chatbot building a ``\textit{dossier}'' of her personal issues. P20 omitted real names or precise timelines: \textit{``I'd be vague... not get exact about timing. You can still be free in what you say, but you gotta be careful how you say it.''}

Four participants (4/21) mentioned sensor-based precautions or platform-specific controls. P1 muted her laptop camera and mic due to a fear of potential surveillance. P6 found an option to disable message retention in ChatGPT, noting, \textit{``I turned that on so it doesn't use my messages to train the model.''} Others supplemented built-in options with third-party tools for added security. P13 explained:
\begin{quote}
\textit{``I use Firefox, I'll do ad block...I'll also use a VPN. I just try to minimize any data leaks within the chatbot itself.''}
\end{quote}
Although such technical measures reflected a stronger awareness of online risks, participants were not always certain of their full effectiveness. P6 asserts that the \textit{``main reason''} she wasn't fully comfortable with ChatGPT for mental health support was because she \textit{``wasn't sure if they [OpenAI] are honest about not using my stuff.''} P13 similarly continued his thoughts on VPN usage with, ``\textit{At the end of the day nothing is one hundred percent, and I'm not sure how a chatbot is identifying me when I use it.}''

\boldpartitle{Eleven participants chose not to employ privacy safeguards, citing trust, futility, or altruism (MC3)}
This group either trusted the AI's manufacturer to behave responsibly or felt that data collection was inevitable. P5 believed, \textit{``No one's really using it...It's just an application I get to express myself with.''} P7 likewise found privacy concerns ``\textit{irrelevant}'' and assumed any collected mental health data ``\textit{could only be used for good.}'' Five participants (5/21) of this eleven who chose not to mitigate expressed outright resignation -- they were convinced that protective efforts were pointless. As P9 conceded, \textit{``I can't stop it anyway... might as well jump on the bandwagon.''} P3 expanded in a similar direction: \textit{``Honestly, anything I share about my mental health is probably sold by whoever is behind ChatGPT, and then used as a way to target my interest, future life decisions, current life decisions, and purchasing power...It's just the way all information goes, really.''}

Within this subset of eleven participants, five (5/21) mentioned that they were motivated to share in an uninhibited capacity by altruism: they saw the open contribution of their mental health struggles as a means to help improve the world. P14 reflected:
\begin{quote}
\textit{``There is a part of me that hopes my data does get used to train AI or whatever... say it helps improve it even the littlest, tiniest bit....maybe I've done my part. Because I don't want other people to go through everything that I've gone through. So if I could improve it a little bit, that's okay.''}
\end{quote}

Seven (7/21) of the eleven acknowledged possible harm but felt the \textit{immediate} emotional relief outweighed distant concerns. As P21 put it, \textit{``Instead of me thinking, `Hey, this AI might become super smart and do something negative in the future,' it was just kind of like, `What can I do right now? How can I help myself right now?''}

\boldpartitle{Opaque privacy policies added confusion and often affected protective behaviors (MC3, MC6)}
Although some participants tried to learn about data handling, many (10/21) found terms-of-service documents cumbersome or uninformative. P2 lamented their length: \textit{``They're usually like 18 pages -- they do it on purpose.''} P16 similarly questioned the transparency of this information: \textit{``I feel like ChatGPT sort of has a way of not being that honest where it's like, `Oh, we use your information,' but it's like, what information do you use? Is it what I'm writing? Is it like the information I logged in with? Like, what do you use exactly? They don't clarify that.''} Only five participants (5/21) expressed an attempted engagement with privacy policies, while more than half (13/21) said that they never read them at all. The resulting uncertainty reinforced privacy resignation for some and contributed to a sense of unease for others; P16 continued her thoughts to share, \textit{``I just hope they're not doing anything really bad... I don't have time to read it, and I probably wouldn't notice anything suspicious anyway.''}

\begin{tcolorbox}[summarybox]
\textbf{``What strategies do users employ (if any) to manage their S\&P concerns when using LLM-enabled conversational agents for mental health support?''}

\textit{Roughly half of our participants attempted to mitigate their concerns by omitting identifiable details such as names and timelines or `cloaking' their experiences in hypothetical or narrative form. About a quarter employed technical privacy-preserving measures, such as using a VPN. Half declined to enact any safeguards at all: justifications included a significant trust in the AI's manufacturer, a sense of futility that their data would be distributed regardless, or altruistic motivations for supplying personal details. The perceived inaccessibility of privacy policies led many to skip or skim available privacy information and continue using LLMs for mental health without taking any privacy-protective actions.}
\end{tcolorbox}

\subsection{Security, Safety, \& Privacy Expectations (RQ3)} \label{sec:results_rq3}

\boldpartitle{Participants experienced in both traditional therapy and LLMs compared privacy and security expectations (MC8, MC9)} We observed varied perspectives on how concerns impacted willingness to disclose personal mental health information to LLM-based tools versus traditional therapists. These perspectives often took the form of salient notes about \textit{safety}, distinct from more technical  security concerns. Participants like P13 tended to expressed a general expectation that their information would be protected to the same capacity as his conversations with a traditional practitioner; for P13 specifically, this was by nature of Gemini being intelligent enough to identify his data as sensitive: 

\begin{quote} ``I definitely would expect [LLMs] to follow the same laws and procedures to protect my health information, because at the end of the day, it’s still sensitive health information. So whether or not you’re doing it through a chatbot, or a counselor, it should all be, you know, under the same umbrella privacy.'' \end{quote} 

Conversely, concerns about data privacy and the potential misuse of information shared with AI tools that would not be present when speaking to therapists made others wary. P19 reflected on the increased risk of data exposure with AI compared to traditional therapy: \textit{“I’m sure it’s just as likely, or maybe even more likely that Gemini or ChatGPT is using that data, more often with greater risk than a therapist... I don't think it has laws really protecting it.”} Only a few participants (3/21) were fully neutral, and generally acknowledged a lack of sufficient awareness regarding legal protections and their applicability to AI chatbots. When prompted if she knew of any legal frameworks concerning mental health data protections, P16 hesitated, before contributing, \textit{``No, I don’t think so. No, I mean, I hope there’s laws and regulations that are like ‘Please do not share her information.’ But no, I don’t know.''}

\boldpartitle{Participants across the board desired clearer privacy practices and dynamic S\&P features (MC5, MC9)}
Regardless of prior experiences with traditional mental health services, most participants voiced frustration with opaque terms-of-service agreements -- thirteen (13/21) explicitly stated that they felt that their conversational agents were currently insufficiently transparent regarding data handling practices. They desired simplified disclosures and adaptive controls. P4 plainly stated he expected xAI to \textit{``Be more on the up and up about where the information goes, how it's stored, where it's gonna be sold to, what's being done with it,''} regarding Grok. He called for \textit{``plain English''} clarity. Nine participants (9/21) wanted LLMs to automatically detect mental health discussions, and offer privacy prompts or ephemeral options. P19 proposed that if the AI detects a user discussing personal struggles, ``\textit{it could pop up…like, `This is private. Do you want to proceed?'}'' P20 expanded on this with a ``\textit{support mode}'' concept that she recognized would be valuable in her mental health conversations with Pi, as a way of offering immediate confidentiality prompts and even the option to delete conversations afterward. In her view, such ephemeral controls were critical because ``\textit{someone in crisis isn't going to stop and think to enable a VPN}.''

\boldpartitle{Other participants called for explicit regulatory or ethical frameworks to protect users (MC5, MC7, MC9)} While acknowledging that tools like ChatGPT and Replika are not built for mental health, five participants (5/21) explicitly wanted legal guidelines preventing misuse of mental health data that might be shared on these platforms. Their perspectives often highlighted the emotional realities faced by vulnerable users. As P20 stressed, ``\textit{I think there needs to be laws and regulations behind that, an eye towards ethics and relatability.}''

Chief among the requested regulatory structure was an expectation for regulators and lawmakers to be more aware of the rapid development of LLMs. P20 continued her thoughts from above to acknowledge her skepticism of the government's current ability to enforce meaningful protections:

\begin{quote}
    ``We have Congress people and senators who literally don’t know how the internet works. They literally don’t know how to turn on the Internet, because these people are fossils.''
\end{quote}

\textit{Protection} itself tended to be perceived as retributive: the five participants who explicitly called for regulatory activity wanted more severe penalties and enforcement mechanisms to deter incidents with their data. P3 was succinct and direct in the expression of her desire for explicit punitive action in the case of any failure to protect her information: \textit{``Protection should look like serious repercussions to those who let data breaches happen, as opposed to just a relatively small million dollar fine when it's a billion dollar company.''}

\boldpartitle{Responsibility for data protection was variably assigned to users, companies, or governments (MC5, MC9)} Six participants (6/21) believed that users themselves held primary responsibility for protecting their data; they emphasized the need for consistent vigilance when disclosing sensitive information. P3 urged individuals to remain cautious. She stated:
\begin{quote} \textit{``The government doesn't care about you. The person behind the AI doesn't care about you because they care about the bottom line, which is money. The only person that will care about you is you.''} \end{quote} 

Eleven participants (11/21) believed that manufacturers are chiefly accountable for data protection given their proximity to the technology and data. These participants stated that the developers had the resources and expertise to implement effective data protection measures, and that users should be able to trust these companies with their sensitive mental health disclosures: as P6 remarked, \textit{``They know what information they're using to train the model...they're responsible for the ethics.''} 

The remaining four (4/21) placed responsibility squarely on the government and explicitly advocated for legislative frameworks. P7 argued that if manufacturers ``\textit{don't have privacy built in, the government needs to implement it.}'' We identify the most significant factor influencing the divergence of responsibility for data protection to be a willingness to adopt proactive mitigatory behaviors -- we explain this effect in greater detail in Section~\ref{sec:discussion_4}.

\begin{tcolorbox}[summarybox]
\textbf{``What are users' expectations regarding data protection and trustworthy interactions when using LLM-enabled conversational agents for mental health support?''}
 
    \textit{Participant expectations were significantly impacted by the degree to which they saw their disclosures as more or less safe than those made to a traditional mental health provider. This variance had a significant role in perceptions of safety. Across these perspectives, users commonly expressed desire for clearer policies and adaptive privacy features. They also diverged in assigning responsibility -- most deferred to manufacturers, some expected individual vigilance, and a few urged formal legislation and regulation on behalf of the government.}
\end{tcolorbox}

\section{Discussion} \label{sec:discussion}

\subsection{The Hidden Costs of AI Chatbots for Mental Health} \label{sec:discussion_1}
We found that user conceptualizations of data privacy were shaped by a gap between how they \emph{thought} their mental health disclosures would be safeguarded, versus the actual technical and legal infrastructures currently governing LLM-based tools. More than half our interviewees identified data privacy as a salient concern, yet most lacked precise knowledge of regulatory or security practices. As such, many did not take any actions to manage their concerns -- and those who did often were not as protected as they might have thought (see Section~\ref{sec:results_rq1}). This echoes scholarship that users often misinterpret or overestimate protective measures~\cite{turow2005open,martin2015privacy} and inadvertently place themselves at risk.

\boldpartitle{Informality and stigma avoidance} Many of our interviewees considered their emotional struggles insufficient to warrant seeking a licensed therapist, and described therapy as both too ``serious'' and potentially stigmatizing -- a perspective that is also aligned with research detailing how many individuals find it difficult to commit to traditional pathways for mental health support~\cite{corrigan2002understanding,vogel2007perceived, sickel2014mental}. Participants described their mental health conversations with LLM chatbots as easily accessible and unstructured spaces, where they vented about everyday concerns or transient emotional distress without feeling that they were engaging with a formal practitioner. This perceived informal and low-barrier environment fostered candid self-disclosures, but also concealed potential risks: we observed that the empathetic and non-judgmental dialogue produced by LLM-enabled conversational agents often enabled interviewees to feel comfortable sharing deeply intimate and affective data (see Section~\ref{sec:results_rq3}).

A similar theme was the consistent reliance on broad, sometimes idealized notions of data safety. Many participants assumed that no unauthorized party would care about their mental health details (see Section~\ref{sec:results_rq1}). Participants also occasionally mapped privacy onto frameworks like HIPAA or doctor-patient confidentiality (see Section~\ref{sec:results_rq1}), without verifying if such regulations applied. This mirrors prior work that highlights misplaced trust when AI systems appear personified or neutral~\cite{lucas2014s,he2023conversational}.

\boldpartitle{\textit{Emergency Privacy Trade-offs}: Urgent mental health needs} Participants overwhelmingly accompanied their thoughts on protection with an acknowledgment that they were largely unaware of how their mental health information was processed, stored, or possibly distributed (see Section~\ref{sec:results_rq1}). Notably, this acknowledgment had little impact on participants' willingness to continue disclosing sensitive details: while the majority of participants were unsure of what was happening to any personally identifiable mental health information they shared, \textit{none} expressed a desire to discontinue using their conversational agent (see Section~\ref{sec:results_rq1}). In urgent mental health scenarios, heightened vulnerability often eclipses privacy concerns: research finds that diminished cognitive resources can lead individuals to accept greater risk for prompt relief~\cite{hindmarch2013depression}. Unlike e-commerce or social media contexts, mental well-being may supersede rational calculations of long-term harm, creating a novel category of ``\textit{emergency privacy trade-offs}'' -- a state of decision-making such that urgent psychological need might take priority. This is also true for general and less extreme cases, as individuals suffering from even moderate anxiety or depression can exhibit decreased information-seeking behavior -- potentially increasing the tendency to accept greater risk for prompt relief~\cite{smith2022lower}.

This surfaced theme aligns with Acquisti and Grossklags's notions of \textit{``bounded rationality''} and \textit{``hyperbolic discounting''}, in which individuals prioritize immediate gratification (i.e., mental health support) over weighing long-term privacy risks~\cite{acquisti2004privacy} in complex cost-benefit scenarios. Several participants felt they had nowhere else to turn for mental health support (see Section~\ref{sec:results_rq1}), and this reinforced a willingness to bypass privacy checks. A pervasive sense of resignation emerged among some who believed it was impossible to fully protect their data (see Section~\ref{sec:results_rq2}), echoing Hoffmann et al.'s concept of ``privacy cynicism''~\cite{hoffmann2016privacy}, wherein users ``discount risks or concerns without ignoring them.''

Participants' limited engagement with privacy-promoting measures was especially concerning given the frequency with which they conveyed \textit{incorrect} understandings of data handling (see Section~\ref{sec:results_rq1}). Privacy scholars consistently identify a knowledge gap between the interactions of U.S.-based consumers with general web-based platforms and legal infrastructures: successive survey studies by Joseph Turow identify that 75\% of people incorrectly believe that a website having a privacy policy means that the site will not share information with other websites and companies~\cite{turow2005open}, and that only 30\% of people correctly answer questions about online privacy~\cite{turow2009americans}. A study by Kirsten Martin demonstrated that people often interpret a privacy notice to be ``more protective of consumer data than [it] actually [is]''~\cite{martin2015privacy}. Martin found that her respondents tended to projected important factors from their own privacy expectations onto available privacy notices, effectively turning these policies into a \emph{tabula rasa} for users' assumptions.

Our interviews revealed a similar effect: many chose not to investigate or alter their behavior in favor of a belief that it was the responsibility of the manufacturer to protect them (see Section~\ref{sec:results_rq3}). Prior work conducted by Hargittai and Marwick finds that, especially among younger users, there is an awareness of privacy risks when disclosing online yet a resignation to limited control~\cite{hargittai2016can}. This combination of \emph{awareness} and \emph{resignation} appeared throughout our interviews, further reinforcing the theme that even those who recognize potential hazards tend to forgo protective behaviors in favor of convenience or emotional relief. Michael Froomkim refers to a similar effect in the context of responses to mass surveillance as ``privacy myopia'', wherein users might find it expensive or impossible to evaluate all long-term ramifications of the transmission of seemingly innocuous personal data~\cite{froomkin2015regulating}. 

\subsection{Intangible Vulnerability} \label{sec:discussion_2}
The concept of \textit{control} becomes especially relevant when considering how differently participants weighted the sensitivity of certain data for privacy and security. Brandimarte et al. describe the increased willingness to disclose when individuals perceive themselves ``in control''~\cite{brandimarte2013misplaced} -- a phenomenon we observed among participants who prided themselves on withholding personal identifiers yet provided enough detail to render them vulnerable. Paradoxically, such attempts at control may embolden sharing that is ultimately risky. The tension between labeling mental health data as both deeply personal yet less vulnerable to immediate harm reveals a nuanced form of misplaced confidence limited to a sense of control, and a misaligned `hierarchy' of information as organized by sensitivity. The most ambitious of our participants attempted to limit or `cloak' detail that they considered to be personally identifiable (see Section~\ref{sec:results_rq2}), specifically to prevent misuse or re-purposing of their mental health disclosures. While resourceful, these attempts are likely to prove insufficient: in multiple interviews (see Sections~\ref{sec:results_rq1} \& ~\ref{sec:results_rq2}), participants described their mental health revelations as sensitive, but considered their sanctity to be less than that of `conventional' sensitive data. This includes financial details such as credit card numbers, and spatiotemporal detail such as home address.

These statements reveal a novel dynamic that we refer to as \textit{``intangible vulnerability''}: participants recognize the personal significance of their disclosures, yet many struggled to envision how non-concrete data, such as anxiety triggers or personal traumas, could be weaponized or monetized. For the selection of interviewees who did note external inferences as a concern, they tended to have varied or incomplete models of explicitly how they figured this would harm them down the line (see Section~\ref{sec:results_rq1}). This phenomenon partially extends Zeynep Tufekci's findings that people imagine concrete harms from data sharing on social networks more readily than intangible repercussions~\cite{tufekci2008can}. Notably, we do this across a new domain by highlighting that emotional or psychological data, while deeply personal, lacks the easily imagined real-world exploit pathways that demand more cautious behavior. Emotional disclosures can be potentially deemed `most private' but paradoxically, `least protected,' because participants fail to map mental health revelations onto the typical frames for data breach or identity theft. \textit{Intangible vulnerability} proves harder to contextualize, and this perpetuates a sense that mental health data is less at risk. 

\subsection{Anthropomorphism, Trust, \& Over-Disclosure} \label{sec:discussion_3}
Anthropomorphism -- the attribution of human-like characteristics to non-human entities -- has been found to play a significant role in fostering user trust and engagement towards technical systems~\cite{spiekermann2001privacy, kim2012anthropomorphism, zlotowski2015anthropomorphism, ischen2020privacy}. As LLMs continue to adopt increasingly human-like communication styles, users may unconsciously attribute mind-like qualities to these systems, further deepening the perceived trust and connection attributed to autonomous systems~\cite{waytz2014mind}. We observed that our participants frequently described their mental health conversations using terms such as \textit{``personalized``} or \textit{''nonjudgmental,``} feelings highlighting the emotional connection promoted by an LLM's anthropomorphic nature (see Section~\ref{sec:results_rq1}). Our interviewees tended to describe feeling comfortable disclosing similar information to their conversational agents that they would to a therapist, often without considering de-identification strategies or recognizing the potential risks of such disclosures (see Section~\ref{sec:results_rq1}). Notably, this aligns with findings that conversational agents can evoke similar emotional, relational, and psychological benefits as disclosures to human partners~\cite{ho2018psychological}.

The stakes of anthropomorphism-induced sharing are especially high when evaluating the usage of conversational agents in mental health contexts, where users are disclosing deeply emotional, sensitive information. Emerging research suggests both that the increased trust in conversational systems as a result of anthropomorphic features is increasingly prevalent in clinical contexts (such as the prescription of medications)~\cite{cohn2024believing}, and that users are increasingly failing to substantially differentiate between the credibility of human-generated and LLM-generated content~\cite{huschens2023you}. As surfaced in our interviews, these findings reflect a broader trend in AI and mental health: users may equate human-like interaction styles with the frameworks in place to protect the disclosures they make to a licensed professionals (see Section~\ref{sec:results_rq1}), despite clear differences in ethical and legal obligations~\cite{marks2023ai}.

\subsection{Perceived Responsibility \& Risk Shape Security \& Privacy Behaviors} \label{sec:discussion_4}
A consistent theme that emerged from our interviews is the interplay between how participants conceptualize who is responsible for safeguarding their mental health disclosures and the privacy and security behaviors they exhibit. Some of our participants expressed the belief that users themselves hold primary responsibility for data protection. Others emphasized the role of manufacturers (or, less commonly, the government) in ensuring confidentiality and protection. These divergent perceptions significantly influenced the degree to which participants engaged in data withholding and other privacy-protective strategies (see Section~\ref{sec:results_rq3}).

\boldpartitle{User-centric responsibility resulted in more protective actions} We observed that users who placed the onus of data protection on themselves frequently exhibited more proactive security and privacy behaviors in the sharing of their mental health details. They described taking deliberate steps to limit identifiable details when discussing sensitive mental health topics with their LLM-enabled tools (see Section~\ref{sec:results_rq2}). When viewed through the lens of Protection Motivation Theory (PMT), an established framework that argues that individuals are more likely to adopt protective behaviors when they perceive a high degree of personal responsibility and vulnerability~\cite{rogers1975protection}, these participants' actions appear consistent with a heightened sense of accountability.

Participants aligning with user-first responsibility echoed sentiments such as, ``\textit{Think of [LLMs] as somebody who's not able to hold a secret.''} They tended to use methods of `cloaking' personal details at higher rates; these same participants often noted that \textit{if} their data were to be leaked, it was up to the user for having shared it -- indicating a strong internalization of potential blame or consequences (see Section~\ref{sec:results_rq3}). Such heightened vigilance is distinct from  ``rule-based'' mental health apps where structured modules might limit opportunities for unguarded and freeform sharing. In LLM chatbots, unstructured conversation can increase risk, but `user-first' participants accept this by building their own guardrails; this phenomenon highlights a user-driven privacy boundary absent in many healthcare-regulated environments. 

\boldpartitle{Other-centric responsibility lead to fewer protective practices} In contrast, participants who viewed manufacturers (or government) as primarily responsible for ensuring data confidentiality tended to be more forthcoming with highly sensitive information. This was traditionally by virtue of the assumption that their information was in good hands; from participants of this mentality, we observed perspectives that their data was not being used in any derogatory capacity (see Sections~\ref{sec:results_rq1} \& ~\ref{sec:results_rq2}). Such trust can be beneficial from an engagement standpoint, as it mirrors the willingness to disclose found in formal therapeutic contexts, but potentially dangerous given the lack of binding regulations for LLM-based tools~\cite{kanter2023health}.

When compared to the findings of existing work assessing user perceptions of security and privacy in technical domains, we observe some similarities. Past research demonstrates that consumers tend to depict security as a responsibility of an arbitrary third party, specifically as a result of a lack of technical ability on their behalf~\cite{haney2021s, gross2007looking, furman2011basing}. For our own participants, we note a similar effect: some who were acutely aware of their unfamiliarity with privacy-preserving measures justified their lack of action or knowledge by indicating that it was the manufacturer's responsibility (see Section~\ref{sec:results_rq3}).

\boldpartitle{The impact of perceived threat} Another dimension that shaped user behavior was \textit{which} entities participants perceived as threatening. Many interviewees identified a single contextual threat between malicious actors (e.g., hackers) or nominally neutral third parties (e.g., employers or insurers). Notably, when participants conceived of ``threat'' in terms of hacking or data breaches, they often stated that they sought to withhold personally identifiable information (see Section~\ref{sec:results_rq2}): this threat vector conjured images of immediate harm (identity theft, doxxing), prompting data minimization or use of alias-based approaches. By contrast, when participants worried about employers or insurers discovering their mental health status, they expressed broader anxiety. This anxiety seldom translated into self-protective strategies. Participants recognized the possibility of adverse consequences such as higher premiums or job discrimination, yet either did not connect these real-world outcomes to the freeform data they shared, or expressed a sense of resignation to whatever may happen (see Section~\ref{sec:results_rq1}).

These contrasting threat models illustrate that privacy-protective behaviors for users of LLM-enabled technologies for mental health are not purely about perceiving risk; they also depend on how users envision potential exploit pathways. The differing influence of privacy standards based on users' perceptions (or perceived threats) is a well-described phenomenon in privacy literature: Helen Nissenbaum's foundational work defining \textit{contextual integrity} emphasizes that privacy concerns arise when information flows deviate from societal norms and expectations within specific contexts~\cite{nissenbaum2004privacy}. 

\subsection{Toward Safer LLMs in Mental Health Support} \label{sec:discussion_5}
A recurrent theme in our data was that participants often advised potential users to read privacy policies and adjust settings -- yet they themselves rarely abide by these same standards (see Section~\ref{sec:results_rq1}). Several interviewees openly acknowledged encouraging peers to be more vigilant and not share too much, yet when asked about their own behaviors, some confessed to skipping policies, and half failed to toggle basic privacy settings: in some cases, these included privacy settings that they reported would make them feel more comfortable, signifying a general lack of awareness (see Section~\ref{sec:results_rq1}). This inconsistency aligns with extensive research documenting mismatches between stated privacy preferences and actual practices. People very rarely read privacy policies~\cite{marotta2011will, milne2004strategies} -- and for understandable reasoning, as McDonald and Cranor famously estimated it would take an impractical amount of time for people to read and understand all of the privacy policies they encounter \cite{mcdonald2008cost}. The vast majority also do not modify privacy settings or opt out of privacy agreements~\cite{janger2001gramm, acquisti2007can}. The tendency of users' stated privacy preferences to not match up with their exhibited privacy behaviors is generally referred to as the ``privacy paradox,'' though experts such as Daniel Solove are notably cautioning this term oversimplifies deeply contextual user behaviors and may obscure the specific reasons behind inconsistent privacy practices~\cite{solove2021myth}.

\boldpartitle{Friction and contextualized privacy concerns} Our findings demonstrate the highly contextual nature of privacy perceptions and risk assessments as motivating mental health disclosures to LLMs (see Sections~\ref{sec:discussion_1} \& ~\ref{sec:discussion_3}). Participants who believe their data to be at risk tend to adjust their sharing behaviors accordingly in a manner that they believe protects them from disadvantageous effects -- this generally looked like the employment of data minimization techniques or alias-based approaches (see Section~\ref{sec:results_rq2}). Others were ambivalent or even optimistic about their mental health data being used for altruistic ends, such as improving AI models or aiding others with similar issues (see Section~\ref{sec:results_rq2}). These attitudes reflect research suggesting that privacy concerns often hinge on the perception of negative downstream uses rather than data distribution \emph{per se}: as posited in a study by Kirsten Martin and Helen Nissenbaum, ``Privacy is not lost, traded off, given away, or violated simply because control over information is ceded or because information is shared or disclosed -- only if ceded or disclosed \emph{inappropriately}''~\cite{martin2016measuring, solove2021myth}. In other words, our participants do not necessarily oppose data sharing in principle; they primarily feared harmful re-contextualization or commercial exploitation.

Further echoing prior studies, we observe that our participants often did not engage with the deeper privacy controls available to them (see Section~\ref{sec:results_rq1}), even though they claim to generally value confidentiality~\cite{acquisti2005privacy, acquisti2006imagined, barth2019putting}. A study by Athey et al. aptly posited that ``whenever privacy requires additional effort or comes at the cost of a less smooth user experience, consumers are quick to abandon technology that would offer them greater protection''~\cite{athey2018c}. This effect is often referred to as \emph{friction}, and is generally used to refer to forces implemented by a company or manufacturer designed to impede or otherwise halt users from the disclosure of personal information in online services~\cite{mcgeveran2013law}. Friction proves especially relevant in mental health contexts, where any added privacy-protective measure can clash with a user's need for empathetic conversation.

\boldpartitle{The role and limits of digital literacy initiatives}  
Alongside design and policy levers, many public and private actors have launched digital literacy programs to help consumers better understand and manage their personal information. Particularly salient is the American Library Association’s (ALA) Media Literacy series, which recently released an episode on demystifying AI in everyday services~\cite{ALA_AI_demystified2024}. Empirical evidence suggests that users' digital literacy and general internet skills have a significant impact on their engagement and relationship with digital privacy \cite{buchi2017caring}. Still, while program availability is expanding, it is not yet dutifully accessible -- a report by the ALA asserted 95 percent of libraries run some form of digital-literacy training, but under 30 percent have digital navigator systems to assist patrons -- and recent federal budget cuts to the Institute of Museum and Library Services threaten key statewide grants~\cite{ALA_report2024, APNews2025}.  
As introduced in our discussion of emergency privacy tradeoffs (see Section~\ref{sec:discussion_1}), users experiencing even minor affective issues may not apply previously learned protective behaviors; similarly, as detailed in this section's discussion on friction, danger is also present in the assumption that users are necessarily going to engage in these digital literacy initiatives even if provided. We assert that while digital literacy interventions can raise baseline awareness if implemented properly, they alone cannot resolve the deep misalignment between mental health needs and privacy self-management burdens in LLM-based settings. 

\boldpartitle{Current legislative gaps: CCPA and developing policy}
We point to the California Consumer Privacy Act (CCPA) as a self-management statute emblematic of structural shortcomings. Although the CCPA allows users to opt out of certain data sales (CIV § 1798.120[b]), it presumes individuals possess the awareness and motivation to protect themselves. Such self-management tacitly permits extensive data use unless users proactively intervene. Our participants typically demonstrated the opposite: minimal follow-through and incomplete mental models demonstrate a consistent desire for frictionless mental health support (see Section~\ref{sec:results_rq1}).

Moreover, while LLM providers including OpenAI and Microsoft explicitly disclaim their tools as \emph{not} mental health services, our study clearly identifies the emergent ways in which users solicit emotional and psychological support from these models. When mental health disclosures occur in general-purpose chats, the risks to privacy and security are especially acute: even if disclaimers distance general-purpose LLMs and their developers from formal `provider' status, the ignoring of this inevitable mental health usage risks invitation of reputational damage and regulatory scrutiny. This is especially prominent given the extent to which these chatbots can do harm to individuals who use them in such ways, as evidenced by a particularly tragic such case involving a Belgian man who died by suicide after engaging in extensive conversations with and finding refuge in an LLM-enabled chatbot~\cite{brusselstimes2023}.We argue that the above tragedy is a very tangible example of the increasingly real danger to people who are using these tools off-label. We thus consolidate our analysis of regulatory gaps and propose a \textit{harm reduction} framework: rather than pushing LLM developers to assume full responsibility as mental health providers, our approach targets data protections supported by realistic legal oversight and minimal user burden.

\boldpartitle{A harm-reduction framework for off-label mental health usage}
We argue for a baseline approach in which regulators and developers adopt proactive measures that protect sensitive user disclosures, designed to be feasible for general-purpose systems and domain-specific systems alike. We identify the following three pillars as paramount:

\textbf{\textit{Contextual nudges \& just-in-time warnings:}} Dynamic and adaptive S\&P responses were the most common participant-requested feature to assuage concerns (see Section~\ref{sec:results_rq3}). As such, we propose that when LLM systems detects language indicative of significant emotional distress or mental health discourse, it should prompt users with a gentle warning: \textit{``We are not a licensed therapist; for confidential crisis support, click here.''} This dynamic intervention aligns with prior research showing that well-placed, just-in-time alerts can foster more cautious user behavior~\cite{athey2018c}. As introduced in our discussion, this recommendation is also in alignment with Nissenbaum's \textit{contextual integrity} (see Section~\ref{sec:discussion_4}). While we observe that some platforms have begun implementing such features, they are not yet standard practice, and are largely inefficient in their current states: a recent study found that while ChatGPT provided highly accurate responses to mental health inquiries, only 33\% included referrals to professional resources~\cite{ayers2023evaluating}. 

\textbf{\textit{Strong default protections and ephemeral storage:}} Most participants never changed default privacy settings, even when doing so could ease their S\&P concerns (see Section~\ref{sec:results_rq3}). As detailed in our discussions, this is further complicated by consistent inabilities to identify difficult to perceive disadvantages to sharing detailed intimate info (see Section~\ref{sec:discussion_2}). We leverage this finding to assert that ephemeral or short-term storage of chat logs should be the default. This measure would especially benefit developers by reducing liability tied to large-scale retention of sensitive data. Users could \emph{opt in} to longer-term storage for convenience, but privacy-by-default avoids placing undue friction on individuals in distress. Practices are currently non-uniform across general-purpose LLMs: ChatGPT retains data for up to 30 days by default, whereas Replika notes that they ``retain your personal information for only as long as necessary to fulfill the purposes we collected it for, including for the purposes of satisfying any legal, accounting, or reporting requirements.''~\cite{replika_privacy_policy}

\textbf{\textit{Targeted oversight and audits:}} Many interviewees assumed companies acted ethically, but were uncertain about any real oversight (see Section~\ref{sec:results_rq2}). As discussed in the prior discussions section, many current regulatory initiatives rely on the attention and commitment of participants (see Section~\ref{sec:discussion_4}). A new or adapted regulatory framework could require third-party audits of data-handling practices, particularly for any product that is \emph{likely} to capture disclosures akin to protected health information. We observe the case of Empatica, a digital mental health technology that has received clearance from the FDA for medical use after years of research and deliberation~\cite{empatica2022}. While we do not suggest that general-purpose LLM-enabled chatbots undergo the same process, a similar degree of rigor should be applied to contractual rules governing data transfers. OpenAI's privacy policies currently indicate that ChatGPT is capable of collecting extensive user data; this includes prompts, responses, email addresses, and geolocation data, which may be shared with affiliates, vendors, service providers, and law enforcement~\cite{openai_privacy_policy}. This lack of transparency raises concerns about where user data may end up -- as surfaced, this is especially critical for sensitive mental health disclosures. We recommend that general-purpose chatbots might instead face a narrower `harm-reduction compliance' requirement that enforces data minimization and transparency of data sharing without placing onus on users.

\boldpartitle{Aligning incentives}
Adopting these harm-reduction strategies not only protects users but can also serve the long-term interests of AI providers. Although disclaimers may reduce immediate liability, they will not shield companies from reputational damage or legal action if leaked or mishandled mental health disclosures spark public outrage or lead to demonstrable harms. As reflected in our participants' attitudes (see Section~\ref{sec:results_rq1}), user trust is tightly coupled with perceptions of data stewardship. Disclaimers currently function as an \emph{escape clause} in theory, but fail to meaningfully curb off-label usage and sensitive disclosure in practice; we predict that both public and policymakers may push for stricter interventions reminiscent of traditional healthcare regulations, as these technologies increasingly pose serious risk to individuals who use them as such \cite{brusselstimes2023}. We advocate for structural changes that embed frictionless privacy features and legally enforceable oversight: such a harm-reduction framework will both protect consumers \emph{and} reduce the liability exposure of AI vendors, to ultimately ensure that all stakeholders benefit from safer and more transparent handling of mental health disclosures in LLM-based systems.

\section{Conclusion} \label{sec:conclusion}

We conducted 21 semi-structured interviews with U.S.-based adopters of LLM-enabled conversational agents for mental health support to better understand their S\&P perspectives. We found that the unrestricted dialogues and uninhibited convenience that make AI-enabled conversational agents appealing also heighten the risk of users inadvertently sharing deeply sensitive information. Our study demonstrated that participants' privacy decisions depend not only on threat models (e.g., bad actors vs. employers) but also on their own perceptions as to what information is ``unsafe''. We introduced the notion of \textit{intangible vulnerability} to capture how deeply personal emotional or psychological disclosures are often undervalued relative to more concrete data such as financial details. This underestimation stems from an inability to envision immediate and tangible harms tied to personal traumas or day-to-day distress, leaving these disclosures comparatively less protected even as they hold powerful insights into a user's mental state. Addressing \textit{intangible vulnerability} thus requires the re-framing of mental health data as equally vulnerable to misuse, even when it lacks the obvious exploit pathways of credit card numbers or addresses. This dynamic demonstrates the urgent need for architectural safeguards and legislative frameworks that account for these intangible harms. 

\section{Ethics Considerations and Open Science} \label{REOS}

\subsection{Ethics Considerations}
Our study protocol was approved by our institution's review board (IRB), and all participants provided informed consent prior to participation. We used Qualtrics, a GDPR-compliant online platform, for the screening survey to identify individuals who had prior experience using LLM-enabled chatbots for mental wellness. Participants who reported no prior use of chatbots for mental wellness were screened out early in the screening survey and received \$0.25 for completing a 1-minute survey. Those who fully completed the 3-minute screening survey received \$0.45. We used Zoom, an end-to-end encrypted (E2EE) platform, for interview sessions. We transcribed, recorded, and stored the data using a secure and password-protected institution-provided system accessible only to the research team. However, we acknowledge that any platforms have their security flaws and hence we try to collect as few identifiers as possible.  Any identifying details mentioned during the interviews (such as names, specific locations, employers, and other personally identifiable information) were removed or replaced with unique identifiers during transcription. Interviews lasted approximately 45 minutes on average, and all interviewees received \$30 via Prolific. We allowed participants the option to keep their cameras off if preferred while the interviewer remained visible on camera. 

We acknowledge that mental health is a deeply sensitive topic, and we gave our participants complete control over what information they chose to share in response to our questions: all interviews were completely self-reported, and participants were given the option to refrain from answering any questions that made them feel uncomfortable in any way. Still, we acknowledge that mental health is a deeply sensitive topic, and discussing it may lead to tangible harm, i.e., psychological exposure to negative memories. To mitigate such harm, we designed an interview script that primarily focused on their experiences with the technology, rather than their mental health conditions. Furthermore, we refrained from getting into deep details about participants’ mental health experiences and conditions to avoid triggering any negative memories.  In addition, one significant part of our interviews revolves around participants’ mitigations, and we believe that these discussions can potentially benefit participants in de-stressing themselves. 

\subsection{Open Science}

Our interview guide and recruitment material are placed in our Available Artifacts. We do not provide full transcripts, so as to prevent our participants from being re-identified by any information they provide that links to their identity. We do not share raw transcripts publicly due to the highly sensitive nature of mental health discussions and to minimize any risk of re-identification. Although the transcripts have been thoroughly anonymized, we remain cautious about the potential for combining contextual or demographic clues that could potentially link any participants to their statements; making these transcripts publicly available could pose tangible risk and violate participant rights to privacy. To promote transparency and reproducibility without compromising confidentiality, we include our interview codebook in the Available Artifacts.

{\balance
{\footnotesize \bibliographystyle{plain}
\bibliography{sample}}}

\begin{thebibliography}{100}

\bibitem{abd2019overview}
Alaa~A Abd-Alrazaq, Mohannad Alajlani, Ali~Abdallah Alalwan, Bridgette~M Bewick, Peter Gardner, and Mowafa Househ.
\newblock An overview of the features of chatbots in mental health: A scoping review.
\newblock {\em International journal of medical informatics}, 132:103978, 2019.

\bibitem{abd2021perceptions}
Alaa~A Abd-Alrazaq, Mohannad Alajlani, Nashva Ali, Kerstin Denecke, Bridgette~M Bewick, and Mowafa Househ.
\newblock Perceptions and opinions of patients about mental health chatbots: scoping review.
\newblock {\em Journal of medical Internet research}, 23(1):e17828, 2021.

\bibitem{MuslimBias}
Abubakar Abid, Maheen Farooqi, and James Zou.
\newblock Large language models associate muslims with violence.
\newblock {\em Nature Machine Intelligence}, 3:461--463, 06 2021.

\bibitem{acquisti2006imagined}
Alessandro Acquisti and Ralph Gross.
\newblock Imagined communities: Awareness, information sharing, and privacy on the facebook.
\newblock In {\em International workshop on privacy enhancing technologies}, pages 36--58. Springer, 2006.

\bibitem{acquisti2004privacy}
Alessandro Acquisti and Jens Grossklags.
\newblock Privacy attitudes and privacy behavior: Losses, gains, and hyperbolic discounting.
\newblock In {\em Economics of information security}, pages 165--178. Springer, 2004.

\bibitem{acquisti2005privacy}
Alessandro Acquisti and Jens Grossklags.
\newblock Privacy and rationality in individual decision making.
\newblock {\em IEEE security \& privacy}, 3(1):26--33, 2005.

\bibitem{acquisti2007can}
Alessandro Acquisti and Jens Grossklags.
\newblock What can behavioral economics teach us about privacy?
\newblock In {\em Digital privacy}, pages 363--378. Auerbach Publications, 2007.

\bibitem{alanezi2024assessing}
Fahad Alanezi.
\newblock Assessing the effectiveness of chatgpt in delivering mental health support: a qualitative study.
\newblock {\em Journal of Multidisciplinary Healthcare}, pages 461--471, 2024.

\bibitem{ALA_report2024}
{American Library Association}.
\newblock New public library technology survey report details digital equity roles, 2024.

\bibitem{ALA_AI_demystified2024}
American~Library Association.
\newblock Media literacy education for adult audiences: Demystifying ai.
\newblock Video, 2024.

\bibitem{athey2018c}
S~Catalini Athey.
\newblock C., \& tucker, ce the digital privacy paradox: Small money, small costs, small talk, 2018.

\bibitem{ayers2023evaluating}
John~W Ayers, Zechariah Zhu, Adam Poliak, Eric~C Leas, Mark Dredze, Michael Hogarth, and Davey~M Smith.
\newblock Evaluating artificial intelligence responses to public health questions.
\newblock {\em JAMA network open}, 6(6):e2317517--e2317517, 2023.

\bibitem{bansalCanAIChatbot}
Bhavika Bansal.
\newblock Can an {{AI Chatbot}} be your therapist?
\newblock https://business.yougov.com/content/49480-can-an-ai-chatbot-be-your-therapist.

\bibitem{barth2019putting}
Susanne Barth, Menno~DT de~Jong, Marianne Junger, Pieter~H Hartel, and Janina~C Roppelt.
\newblock Putting the privacy paradox to the test: Online privacy and security behaviors among users with technical knowledge, privacy awareness, and financial resources.
\newblock {\em Telematics and informatics}, 41:55--69, 2019.

\bibitem{blease2024psychiatrists}
Charlotte Blease, Abigail Worthen, and John Torous.
\newblock Psychiatrists’ experiences and opinions of generative artificial intelligence in mental healthcare: An online mixed methods survey.
\newblock {\em Psychiatry Research}, 333:115724, 2024.

\bibitem{brandimarte2013misplaced}
Laura Brandimarte, Alessandro Acquisti, and George Loewenstein.
\newblock Misplaced confidences: Privacy and the control paradox.
\newblock {\em Social psychological and personality science}, 4(3):340--347, 2013.

\bibitem{buchi2017caring}
Moritz B{\"u}chi, Natascha Just, and Michael Latzer.
\newblock Caring is not enough: The importance of internet skills for online privacy protection.
\newblock {\em Information, Communication \& Society}, 20(8):1261--1278, 2017.

\bibitem{burton2016pilot}
Christopher Burton, Aurora Szentagotai~Tatar, Brian McKinstry, Colin Matheson, Silviu Matu, Ramona Moldovan, Michele Macnab, Elaine Farrow, Daniel David, Claudia Pagliari, et~al.
\newblock Pilot randomised controlled trial of help4mood, an embodied virtual agent-based system to support treatment of depression.
\newblock {\em Journal of telemedicine and telecare}, 22(6):348--355, 2016.

\bibitem{cameron2019assessing}
Gillian Cameron, David Cameron, Gavin Megaw, Raymond Bond, Maurice Mulvenna, Siobhan O’Neill, Cherie Armour, and Michael McTear.
\newblock Assessing the usability of a chatbot for mental health care.
\newblock In {\em Internet Science: INSCI 2018 International Workshops, St. Petersburg, Russia, October 24--26, 2018, Revised Selected Papers 5}, pages 121--132. Springer, 2019.

\bibitem{cdc_wonder}
{Centers for Disease Control and Prevention}.
\newblock Cdc wonder - multiple cause of death, 1999-2022, 2022.

\bibitem{chen2023can}
Yang Chen, Ethan Mendes, Sauvik Das, Wei Xu, and Alan Ritter.
\newblock Can language models be instructed to protect personal information?
\newblock {\em arXiv preprint arXiv:2310.02224}, 2023.

\bibitem{clement2015impact}
Sarah Clement, Oliver Schauman, Tanya Graham, Francesca Maggioni, Sara Evans-Lacko, Nikita Bezborodovs, Craig Morgan, Nicolas R{\"u}sch, June~SL Brown, and Graham Thornicroft.
\newblock What is the impact of mental health-related stigma on help-seeking? a systematic review of quantitative and qualitative studies.
\newblock {\em Psychological medicine}, 45(1):11--27, 2015.

\bibitem{cohn2024believing}
Michelle Cohn, Mahima Pushkarna, Gbolahan~O Olanubi, Joseph~M Moran, Daniel Padgett, Zion Mengesha, and Courtney Heldreth.
\newblock Believing anthropomorphism: Examining the role of anthropomorphic cues on trust in large language models.
\newblock In {\em Extended Abstracts of the CHI Conference on Human Factors in Computing Systems}, pages 1--15, 2024.

\bibitem{corbin2015basics}
Juliet Corbin and Anselm Strauss.
\newblock {\em Basics of qualitative research}, volume~14.
\newblock sage, 2015.

\bibitem{corrigan2004stigma}
Patrick Corrigan.
\newblock How stigma interferes with mental health care.
\newblock {\em American psychologist}, 59(7):614, 2004.

\bibitem{corrigan2002understanding}
Patrick~W Corrigan and Amy~C Watson.
\newblock Understanding the impact of stigma on people with mental illness.
\newblock {\em World psychiatry}, 1(1):16, 2002.

\bibitem{de2023benefits}
Munmun De~Choudhury, Sachin~R Pendse, and Neha Kumar.
\newblock Benefits and harms of large language models in digital mental health.
\newblock {\em arXiv preprint arXiv:2311.14693}, 2023.

\bibitem{de2024health}
Julian De~Freitas and I~Glenn Cohen.
\newblock The health risks of generative ai-based wellness apps.
\newblock {\em Nature Medicine}, pages 1--7, 2024.

\bibitem{Elyoseph2024AssessingPI}
Zohar Elyoseph, Inbar Levkovich, and Shiri Shinan-Altman.
\newblock Assessing prognosis in depression: comparing perspectives of ai models, mental health professionals and the general public.
\newblock {\em Family Medicine and Community Health}, 12, 2024.

\bibitem{empatica2022}
Empatica.
\newblock The empatica health monitoring platform receives fda clearance, 2022.

\bibitem{fitzpatrick2017delivering}
Kathleen~Kara Fitzpatrick, Alison Darcy, and Molly Vierhile.
\newblock Delivering cognitive behavior therapy to young adults with symptoms of depression and anxiety using a fully automated conversational agent (woebot): a randomized controlled trial.
\newblock {\em JMIR mental health}, 4(2):e7785, 2017.

\bibitem{guidancegeneral}
Center for Devices, US~Food Radiological~health, and Drug Admin.
\newblock General wellness: Policy for low risk devices.

\bibitem{froomkin2015regulating}
A~Michael Froomkin.
\newblock Regulating mass surveillance as privacy pollution: Learning from environemntal impact statements.
\newblock {\em U. Ill. L. Rev.}, page 1713, 2015.

\bibitem{fulmer2018using}
Russell Fulmer, Angela Joerin, Breanna Gentile, Lysanne Lakerink, Michiel Rauws, et~al.
\newblock Using psychological artificial intelligence (tess) to relieve symptoms of depression and anxiety: randomized controlled trial.
\newblock {\em JMIR mental health}, 5(4):e9782, 2018.

\bibitem{fung2006early}
Vicki Fung, Eduardo Ortiz, Jie Huang, Bruce Fireman, Robert Miller, Joseph~V Selby, and John Hsu.
\newblock Early experiences with e-health services (1999--2002): promise, reality, and implications.
\newblock {\em Medical care}, 44(5):491--496, 2006.

\bibitem{furman2011basing}
Susanne Furman, Mary~Frances Theofanos, Yee-Yin Choong, and Brian Stanton.
\newblock Basing cybersecurity training on user perceptions.
\newblock {\em IEEE Security \& Privacy}, 10(2):40--49, 2011.

\bibitem{gao2023retrieval}
Yunfan Gao, Yun Xiong, Xinyu Gao, Kangxiang Jia, Jinliu Pan, Yuxi Bi, Yi~Dai, Jiawei Sun, and Haofen Wang.
\newblock Retrieval-augmented generation for large language models: A survey.
\newblock {\em arXiv preprint arXiv:2312.10997}, 2023.

\bibitem{gross2007looking}
Joshua~B Gross and Mary~Beth Rosson.
\newblock Looking for trouble: understanding end-user security management.
\newblock In {\em Proceedings of the 2007 Symposium on Computer Human interaction For the Management of information Technology}, pages 10--es, 2007.

\bibitem{haney2021s}
Julie Haney, Yasemin Acar, and Susanne Furman.
\newblock " it's the company, the government, you and i": User perceptions of responsibility for smart home privacy and security.
\newblock In {\em 30th USENIX Security Symposium (USENIX Security 21)}, pages 411--428, 2021.

\bibitem{hargittai2016can}
Eszter Hargittai and Alice Marwick.
\newblock “what can i really do?” explaining the privacy paradox with online apathy.
\newblock {\em International journal of communication}, 10:21, 2016.

\bibitem{harrer2023attention}
Stefan Harrer.
\newblock Attention is not all you need: the complicated case of ethically using large language models in healthcare and medicine.
\newblock {\em EBioMedicine}, 90, 2023.

\bibitem{he2024physician}
Wenjie He, Wenyan Zhang, Ya~Jin, Qiang Zhou, Huadan Zhang, and Qing Xia.
\newblock Physician versus large language model chatbot responses to web-based questions from autistic patients in chinese: Cross-sectional comparative analysis.
\newblock {\em Journal of Medical Internet Research}, 26:e54706, 2024.

\bibitem{he2023conversational}
Yuhao He, Li~Yang, Chunlian Qian, Tong Li, Zhengyuan Su, Qiang Zhang, and Xiangqing Hou.
\newblock Conversational agent interventions for mental health problems: systematic review and meta-analysis of randomized controlled trials.
\newblock {\em Journal of medical Internet research}, 25:e43862, 2023.

\bibitem{woebot2021survey}
Woebot Health.
\newblock A paradigm shift: Consumer attitudes toward mental health technology in 2021, 2021.

\bibitem{hrsa2025}
{Health Resources and Services Administration}.
\newblock Designated health professional shortage areas statistics.
\newblock \url{https://data.hrsa.gov/topics/health-workforce/shortage-areas}, 2025.

\bibitem{hindmarch2013depression}
Thomas Hindmarch, Matthew Hotopf, and Gareth~S Owen.
\newblock Depression and decision-making capacity for treatment or research: a systematic review.
\newblock {\em BMC medical ethics}, 14:1--10, 2013.

\bibitem{ho2018psychological}
Annabell Ho, Jeff Hancock, and Adam~S Miner.
\newblock Psychological, relational, and emotional effects of self-disclosure after conversations with a chatbot.
\newblock {\em Journal of Communication}, 68(4):712--733, 2018.

\bibitem{hoffmann2016privacy}
Christian~Pieter Hoffmann, Christoph Lutz, and Giulia Ranzini.
\newblock Privacy cynicism: A new approach to the privacy paradox.
\newblock {\em Cyberpsychology: Journal of Psychosocial Research on Cyberspace}, 10(4), 2016.

\bibitem{huang2020challenges}
Minlie Huang, Xiaoyan Zhu, and Jianfeng Gao.
\newblock Challenges in building intelligent open-domain dialog systems.
\newblock {\em ACM Transactions on Information Systems (TOIS)}, 38(3):1--32, 2020.

\bibitem{huschens2023you}
Martin Huschens, Martin Briesch, Dominik Sobania, and Franz Rothlauf.
\newblock Do you trust chatgpt?--perceived credibility of human and ai-generated content.
\newblock {\em arXiv preprint arXiv:2309.02524}, 2023.

\bibitem{huygens2015internet}
Martine~WJ Huygens, Joan Vermeulen, Roland~D Friele, Onno~CP van Schayck, Judith~D de~Jong, and Luc~P de~Witte.
\newblock Internet services for communicating with the general practice: barely noticed and used by patients.
\newblock {\em Interactive journal of medical research}, 4(4):e4245, 2015.

\bibitem{inkster2018empathy}
Becky Inkster, Shubhankar Sarda, Vinod Subramanian, et~al.
\newblock An empathy-driven, conversational artificial intelligence agent (wysa) for digital mental well-being: real-world data evaluation mixed-methods study.
\newblock {\em JMIR mHealth and uHealth}, 6(11):e12106, 2018.

\bibitem{ischen2020privacy}
Carolin Ischen, Theo Araujo, Hilde Voorveld, Guda van Noort, and Edith Smit.
\newblock Privacy concerns in chatbot interactions.
\newblock In {\em Chatbot Research and Design: Third International Workshop, CONVERSATIONS 2019, Amsterdam, The Netherlands, November 19--20, 2019, Revised Selected Papers 3}, pages 34--48. Springer, 2020.

\bibitem{janger2001gramm}
Edward~J Janger and Paul~M Schwartz.
\newblock The gramm-leach-bliley act, information privacy, and the limits of default rules.
\newblock {\em Minn. L. Rev.}, 86:1219, 2001.

\bibitem{TessWSJ}
Julie Jargon.
\newblock A chatbot was designed to help prevent eating disorders. then it gave dieting tips.
\newblock {\em The Wall Street Journal}, 2023.

\bibitem{kanter2023health}
Genevieve~P Kanter and Eric~A Packel.
\newblock Health care privacy risks of ai chatbots.
\newblock {\em JAMA}, 2023.

\bibitem{kasneci2023chatgpt}
Enkelejda Kasneci, Kathrin Se{\ss}ler, Stefan K{\"u}chemann, Maria Bannert, Daryna Dementieva, Frank Fischer, Urs Gasser, Georg Groh, Stephan G{\"u}nnemann, Eyke H{\"u}llermeier, et~al.
\newblock Chatgpt for good? on opportunities and challenges of large language models for education.
\newblock {\em Learning and individual differences}, 103:102274, 2023.

\bibitem{kim2024mindful}
Taewan Kim, Seolyeong Bae, Hyun~Ah Kim, Su-Woo Lee, Hwajung Hong, Chanmo Yang, and Young-Ho Kim.
\newblock Mindfuldiary: Harnessing large language model to support psychiatric patients' journaling.
\newblock In {\em Proceedings of the 2024 CHI Conference on Human Factors in Computing Systems}, CHI '24, New York, NY, USA, 2024. Association for Computing Machinery.

\bibitem{kim2012anthropomorphism}
Youjeong Kim and S~Shyam Sundar.
\newblock Anthropomorphism of computers: Is it mindful or mindless?
\newblock {\em Computers in Human Behavior}, 28(1):241--250, 2012.

\bibitem{koutsouleris2022promise}
Nikolaos Koutsouleris, Tobias~U Hauser, Vasilisa Skvortsova, and Munmun De~Choudhury.
\newblock From promise to practice: towards the realisation of ai-informed mental health care.
\newblock {\em The Lancet Digital Health}, 4(11):e829--e840, 2022.

\bibitem{kuehn2022clinician}
Bridget~M Kuehn.
\newblock Clinician shortage exacerbates pandemic-fueled “mental health crisis”.
\newblock {\em JAMA}, 327(22):2179--2181, 2022.

\bibitem{lawrence2024opportunities}
Hannah~R Lawrence, Renee~A Schneider, Susan~B Rubin, Maja~J Matari{\'c}, Daniel~J McDuff, and Megan~Jones Bell.
\newblock The opportunities and risks of large language models in mental health.
\newblock {\em JMIR Mental Health}, 11(1):e59479, 2024.

\bibitem{lee2020designing}
Yi-Chieh Lee, Naomi Yamashita, and Yun Huang.
\newblock Designing a chatbot as a mediator for promoting deep self-disclosure to a real mental health professional.
\newblock {\em Proceedings of the ACM on Human-Computer Interaction}, 4(CSCW1):1--27, 2020.

\bibitem{lee2020hear}
Yi-Chieh Lee, Naomi Yamashita, Yun Huang, and Wai Fu.
\newblock " i hear you, i feel you": encouraging deep self-disclosure through a chatbot.
\newblock In {\em Proceedings of the 2020 CHI conference on human factors in computing systems}, pages 1--12, 2020.

\bibitem{legaspi2022user}
Carlos~Miguel Legaspi~Jr, Tristan~Raphael Pacana, Kyle Loja, Christina Sing, and Ethel Ong.
\newblock User perception of wysa as a mental well-being support tool during the covid-19 pandemic.
\newblock In {\em Proceedings of the Asian HCI Symposium 2022}, pages 52--57, 2022.

\bibitem{li2023systematic}
Han Li, Renwen Zhang, Yi-Chieh Lee, Robert~E Kraut, and David~C Mohr.
\newblock Systematic review and meta-analysis of ai-based conversational agents for promoting mental health and well-being.
\newblock {\em NPJ Digital Medicine}, 6(1):236, 2023.

\bibitem{loh2023harnessing}
Siyuan~Brandon Loh and Aravind~Sesagiri Raamkumar.
\newblock Harnessing large language models' empathetic response generation capabilities for online mental health counselling support.
\newblock {\em arXiv preprint arXiv:2310.08017}, 2023.

\bibitem{lucas2014s}
Gale~M Lucas, Jonathan Gratch, Aisha King, and Louis-Philippe Morency.
\newblock It’s only a computer: Virtual humans increase willingness to disclose.
\newblock {\em Computers in Human Behavior}, 37:94--100, 2014.

\bibitem{ly2017fully}
Kien~Hoa Ly, Ann-Marie Ly, and Gerhard Andersson.
\newblock A fully automated conversational agent for promoting mental well-being: a pilot rct using mixed methods.
\newblock {\em Internet interventions}, 10:39--46, 2017.

\bibitem{ma2023understanding}
Zilin Ma, Yiyang Mei, and Zhaoyuan Su.
\newblock Understanding the benefits and challenges of using large language model-based conversational agents for mental well-being support.
\newblock In {\em AMIA Annual Symposium Proceedings}, volume 2023, page 1105. American Medical Informatics Association, 2023.

\bibitem{malik2022evaluating}
Tanya Malik, Adrian~Jacques Ambrose, Chaitali Sinha, et~al.
\newblock Evaluating user feedback for an artificial intelligence--enabled, cognitive behavioral therapy--based mental health app (wysa): qualitative thematic analysis.
\newblock {\em JMIR Human Factors}, 9(2):e35668, 2022.

\bibitem{marks2023ai}
Mason Marks and Claudia~E Haupt.
\newblock Ai chatbots, health privacy, and challenges to hipaa compliance.
\newblock {\em Jama}, 2023.

\bibitem{marotta2011will}
Florencia Marotta-Wurgler.
\newblock Will increased disclosure help-evaluating the recommendations of the ali's principles of the law of software contracts.
\newblock {\em U. Chi. L. Rev.}, 78:165, 2011.

\bibitem{martin2015privacy}
Kirsten Martin.
\newblock Privacy notices as tabula rasa: An empirical investigation into how complying with a privacy notice is related to meeting privacy expectations online.
\newblock {\em Journal of Public Policy \& Marketing}, 34(2):210--227, 2015.

\bibitem{martin2016measuring}
Kirsten Martin and Helen Nissenbaum.
\newblock Measuring privacy: An empirical test using context to expose confounding variables.
\newblock {\em Colum. Sci. \& Tech. L. Rev.}, 18:176, 2016.

\bibitem{mcdonald2008cost}
Aleecia~M McDonald and Lorrie~Faith Cranor.
\newblock The cost of reading privacy policies.
\newblock {\em Isjlp}, 4:543, 2008.

\bibitem{mcgeveran2013law}
William McGeveran.
\newblock The law of friction.
\newblock {\em U. Chi. Legal F.}, page~15, 2013.

\bibitem{mhanational2024}
{Mental Health America}.
\newblock Mental health america: Adult data 2024, 2024.

\bibitem{milne2004strategies}
George~R Milne and Mary~J Culnan.
\newblock Strategies for reducing online privacy risks: Why consumers read (or don’t read) online privacy notices.
\newblock {\em Journal of interactive marketing}, 18(3):15--29, 2004.

\bibitem{mongelli2020challenges}
Francesca Mongelli, Penelope Georgakopoulos, and Michele~T Pato.
\newblock Challenges and opportunities to meet the mental health needs of underserved and disenfranchised populations in the united states.
\newblock {\em Focus}, 18(1):16--24, 2020.

\bibitem{APNews2025}
{Nadia Lathan}.
\newblock Institute of museum and library services funding cuts under trump order, 2025.

\bibitem{nissenbaum2004privacy}
Helen Nissenbaum.
\newblock Privacy as contextual integrity.
\newblock {\em Wash. L. Rev.}, 79:119, 2004.

\bibitem{olano2022effectiveness}
Eduardo Olano-Espinosa, Jose~Francisco Avila-Tomas, Cesar Minue-Lorenzo, Blanca Matilla-Pardo, Mar{\'\i}a Encarnaci{\'o}n~Serrano Serrano, F~Javier Martinez-Suberviola, Mario Gil-Conesa, Isabel Del Cura-Gonz{\'a}lez, et~al.
\newblock Effectiveness of a conversational chatbot (dejal@ bot) for the adult population to quit smoking: pragmatic, multicenter, controlled, randomized clinical trial in primary care.
\newblock {\em JMIR mHealth and uHealth}, 10(6):e34273, 2022.

\bibitem{Omiye2023.07.03.23292192}
Jesutofunmi~A. Omiye, Jenna Lester, Simon Spichak, Veronica Rotemberg, and Roxana Daneshjou.
\newblock Beyond the hype: large language models propagate race-based medicine.
\newblock {\em medRxiv}, 2023.

\bibitem{openai_privacy_policy}
{OpenAI}.
\newblock Privacy policy, 2024.

\bibitem{orne2009demand}
Martin~T Orne.
\newblock Demand characteristics and the concept of quasi-controls.
\newblock {\em Artifacts in behavioral research: Robert Rosenthal and Ralph L. Rosnow’s classic books}, 110:110--137, 2009.

\bibitem{o2022massive}
Daniel~E O’Leary.
\newblock Massive data language models and conversational artificial intelligence: Emerging issues.
\newblock {\em Intelligent Systems in Accounting, Finance and Management}, 29(3):182--198, 2022.

\bibitem{palmer2022beneficent}
Amitabha Palmer and David Schwan.
\newblock Beneficent dehumanization: Employing artificial intelligence and carebots to mitigate shame-induced barriers to medical care.
\newblock {\em Bioethics}, 36(2):187--193, 2022.

\bibitem{panch2019artificial}
Trishan Panch, Heather Mattie, and Rifat Atun.
\newblock Artificial intelligence and algorithmic bias: implications for health systems.
\newblock {\em Journal of global health}, 9(2), 2019.

\bibitem{park2021wrote}
SoHyun Park, Anja Thieme, Jeongyun Han, Sungwoo Lee, Wonjong Rhee, and Bongwon Suh.
\newblock “i wrote as if i were telling a story to someone i knew.”: Designing chatbot interactions for expressive writing in mental health.
\newblock In {\em Proceedings of the 2021 ACM Designing Interactive Systems Conference}, pages 926--941, 2021.

\bibitem{Perlis2024ClinicalDS}
Roy~H. Perlis, Joseph~F Goldberg, Michael~Joshua Ostacher, and Christopher~D Schneck.
\newblock Clinical decision support for bipolar depression using large language models.
\newblock {\em Neuropsychopharmacology}, 49:1412 -- 1416, 2024.

\bibitem{prochaska2021therapeutic}
Judith~J Prochaska, Erin~A Vogel, Amy Chieng, Matthew Kendra, Michael Baiocchi, Sarah Pajarito, and Athena Robinson.
\newblock A therapeutic relational agent for reducing problematic substance use (woebot): development and usability study.
\newblock {\em Journal of medical Internet research}, 23(3):e24850, 2021.

\bibitem{replika_privacy_policy}
{Replika}.
\newblock Privacy policy, 2024.

\bibitem{rogers1975protection}
Ronald~W Rogers.
\newblock A protection motivation theory of fear appeals and attitude change1.
\newblock {\em The journal of psychology}, 91(1):93--114, 1975.

\bibitem{JS21}
Johnny Saldaña.
\newblock {\em The coding manual for qualitative researchers}.
\newblock sage, 2021.

\bibitem{schroeder2018pocket}
Jessica Schroeder, Chelsey Wilkes, Kael Rowan, Arturo Toledo, Ann Paradiso, Mary Czerwinski, Gloria Mark, and Marsha~M Linehan.
\newblock Pocket skills: A conversational mobile web app to support dialectical behavioral therapy.
\newblock In {\em Proceedings of the 2018 CHI Conference on Human Factors in Computing Systems}, pages 1--15, 2018.

\bibitem{scrutton2017epistemic}
Anastasia~Philippa Scrutton.
\newblock Epistemic injustice and mental illness.
\newblock In {\em The Routledge handbook of epistemic injustice}, pages 347--355. Routledge, 2017.

\bibitem{sezgin2023clinical}
Emre Sezgin, Faraaz Chekeni, Jennifer Lee, and Sarah Keim.
\newblock Clinical accuracy of large language models and google search responses to postpartum depression questions: cross-sectional study.
\newblock {\em Journal of Medical Internet Research}, 25:e49240, 2023.

\bibitem{sharma2023cognitive}
Ashish Sharma, Kevin Rushton, Inna~Wanyin Lin, David Wadden, Khendra~G Lucas, Adam~S Miner, Theresa Nguyen, and Tim Althoff.
\newblock Cognitive reframing of negative thoughts through human-language model interaction.
\newblock {\em arXiv preprint arXiv:2305.02466}, 2023.

\bibitem{sickel2014mental}
Amy~E Sickel, Jason~D Seacat, and Nina~A Nabors.
\newblock Mental health stigma update: A review of consequences.
\newblock {\em Advances in Mental Health}, 12(3):202--215, 2014.

\bibitem{siddals2024just}
Steven Siddals, Astrid Coxon, and John Torous.
\newblock " it just happened to be the perfect thing": Real-life experiences of generative ai chatbots for mental health.
\newblock 2024.

\bibitem{simon2022skating}
David~A Simon, Carmel Shachar, and I~Glenn Cohen.
\newblock Skating the line between general wellness products and regulated devices: strategies and implications.
\newblock {\em Journal of Law and the Biosciences}, 9(2):lsac015, 2022.

\bibitem{smith2014virtual}
Matthew~J Smith, Emily~J Ginger, Katherine Wright, Michael~A Wright, Julie~Lounds Taylor, Laura~Boteler Humm, Dale~E Olsen, Morris~D Bell, and Michael~F Fleming.
\newblock Virtual reality job interview training in adults with autism spectrum disorder.
\newblock {\em Journal of autism and developmental disorders}, 44:2450--2463, 2014.

\bibitem{smith2022lower}
Ryan Smith, Samuel Taylor, Robert~C Wilson, Anne~E Chuning, Michelle~R Persich, Siyu Wang, and William~DS Killgore.
\newblock Lower levels of directed exploration and reflective thinking are associated with greater anxiety and depression.
\newblock {\em Frontiers in Psychiatry}, 12:782136, 2022.

\bibitem{so2024guided}
Ryuhei So, Naoki Emura, Kozue Okazaki, Sakiko Takeda, Takashi Sunami, Kohei Kitagawa, Yoshitake Takebayashi, and Toshi~A Furukawa.
\newblock Guided versus unguided chatbot-delivered cognitive behavioral intervention for individuals with moderate-risk and problem gambling: A randomized controlled trial (gambot2 study).
\newblock {\em Addictive behaviors}, 149:107889, 2024.

\bibitem{solove2021myth}
Daniel~J Solove.
\newblock The myth of the privacy paradox.
\newblock {\em Geo. Wash. L. Rev.}, 89:1, 2021.

\bibitem{songTypingCureExperiences2024}
Inhwa Song, Sachin~R. Pendse, Neha Kumar, and Munmun De~Choudhury.
\newblock The {{Typing Cure}}: {{Experiences}} with {{Large Language Model Chatbots}} for {{Mental Health Support}}, March 2024.

\bibitem{spiekermann2001privacy}
Sarah Spiekermann, Jens Grossklags, and Bettina Berendt.
\newblock E-privacy in 2nd generation e-commerce: privacy preferences versus actual behavior.
\newblock In {\em Proceedings of the 3rd ACM conference on Electronic Commerce}, pages 38--47, 2001.

\bibitem{ta2020user}
Vivian Ta, Caroline Griffith, Carolynn Boatfield, Xinyu Wang, Maria Civitello, Haley Bader, Esther DeCero, Alexia Loggarakis, et~al.
\newblock User experiences of social support from companion chatbots in everyday contexts: thematic analysis.
\newblock {\em Journal of medical Internet research}, 22(3):e16235, 2020.

\bibitem{thomas2006general}
David~R Thomas.
\newblock A general inductive approach for analyzing qualitative evaluation data.
\newblock {\em American journal of evaluation}, 27(2):237--246, 2006.

\bibitem{brusselstimes2023}
The~Brussels Times.
\newblock Belgian man commits suicide following exchanges with chatbot, 2023.

\bibitem{tufekci2008can}
Zeynep Tufekci.
\newblock Can you see me now? audience and disclosure regulation in online social network sites.
\newblock {\em Bulletin of Science, Technology \& Society}, 28(1):20--36, 2008.

\bibitem{turow2005open}
Joseph Turow, Lauren Feldman, and Kimberly Meltzer.
\newblock Open to exploitation: American shoppers online and offline, 2005.

\bibitem{turow2009americans}
Joseph Turow, Jennifer King, Chris~Jay Hoofnagle, Amy Bleakley, and Michael Hennessy.
\newblock Americans reject tailored advertising and three activities that enable it.
\newblock {\em Available at SSRN 1478214}, 2009.

\bibitem{vaidyam2019chatbots}
Aditya~Nrusimha Vaidyam, Hannah Wisniewski, John~David Halamka, Matcheri~S Kashavan, and John~Blake Torous.
\newblock Chatbots and conversational agents in mental health: a review of the psychiatric landscape.
\newblock {\em The Canadian Journal of Psychiatry}, 64(7):456--464, 2019.

\bibitem{vogel2007perceived}
David~L Vogel, Nathaniel~G Wade, and Ashley~H Hackler.
\newblock Perceived public stigma and the willingness to seek counseling: The mediating roles of self-stigma and attitudes toward counseling.
\newblock {\em Journal of counseling psychology}, 54(1):40, 2007.

\bibitem{waytz2014mind}
Adam Waytz, Joy Heafner, and Nicholas Epley.
\newblock The mind in the machine: Anthropomorphism increases trust in an autonomous vehicle.
\newblock {\em Journal of experimental social psychology}, 52:113--117, 2014.

\bibitem{weidinger2021ethical}
Laura Weidinger, John Mellor, Maribeth Rauh, Conor Griffin, Jonathan Uesato, Po-Sen Huang, Myra Cheng, Mia Glaese, Borja Balle, Atoosa Kasirzadeh, et~al.
\newblock Ethical and social risks of harm from language models.
\newblock {\em arXiv preprint arXiv:2112.04359}, 2021.

\bibitem{wendler2006racial}
David Wendler, Raynard Kington, Jennifer Madans, Gretchen~Van Wye, Heidi Christ-Schmidt, Laura~A Pratt, Otis~W Brawley, Cary~P Gross, and Ezekiel Emanuel.
\newblock Are racial and ethnic minorities less willing to participate in health research?
\newblock {\em PLoS medicine}, 3(2):e19, 2006.

\bibitem{wornow2023shaky}
Michael Wornow, Yizhe Xu, Rahul Thapa, Birju Patel, Ethan Steinberg, Scott Fleming, Michael~A Pfeffer, Jason Fries, and Nigam~H Shah.
\newblock The shaky foundations of large language models and foundation models for electronic health records.
\newblock {\em npj Digital Medicine}, 6(1):135, 2023.

\bibitem{ElizaVice}
Chloe Xiang.
\newblock ‘he would still be here’: Man dies by suicide after talking with ai chatbot, widow says.
\newblock {\em Vice}, 2023.

\bibitem{Zack2023.07.13.23292577}
Travis Zack, Eric Lehman, Mirac Suzgun, Jorge~A. Rodriguez, Leo~Anthony Celi, Judy Gichoya, Dan Jurafsky, Peter Szolovits, David~W. Bates, Raja-Elie~E. Abdulnour, Atul~J. Butte, and Emily Alsentzer.
\newblock Coding inequity: Assessing gpt-4{\textquoteright}s potential for perpetuating racial and gender biases in healthcare.
\newblock {\em medRxiv}, 2023.

\bibitem{zagorskiPopularityMentalHealth2022}
Nick Zagorski.
\newblock Popularity of {{Mental Health Chatbots Grows}}.
\newblock {\em Psychiatric News}, 57(5), May 2022.

\bibitem{zhong2023artificial}
Yi~Zhong, Yu-jun Chen, Yang Zhou, Jia-Jun Yin, Yu-jun Gao, et~al.
\newblock The artificial intelligence large language models and neuropsychiatry practice and research ethic.
\newblock {\em Asian journal of psychiatry}, 84:103577, 2023.

\bibitem{ziems2022inducing}
Caleb Ziems, Minzhi Li, Anthony Zhang, and Diyi Yang.
\newblock Inducing positive perspectives with text reframing.
\newblock {\em arXiv preprint arXiv:2204.02952}, 2022.

\bibitem{zlotowski2015anthropomorphism}
Jakub Z{\l}otowski, Diane Proudfoot, Kumar Yogeeswaran, and Christoph Bartneck.
\newblock Anthropomorphism: opportunities and challenges in human--robot interaction.
\newblock {\em International journal of social robotics}, 7:347--360, 2015.

\end{thebibliography}

\appendix

\section*{Available Artifacts} \label{sec:artifacts}

Attached to the following link we make available all artifacts relevant to our study and promised in our `Open Science' section. These include interview protocol, screening surveys, informed consent, and full codebook: https://doi.org/10.5281/zenodo.15596321

\end{document}